\documentclass[final]{siamltex} 
\usepackage{amsmath} 
\usepackage{amssymb} 
\usepackage{caption}
\usepackage{delarray} 
\usepackage{dsfont} 
\usepackage{epsfig} 
\usepackage{multirow} 
\usepackage{subfigure} 
\usepackage[usenames]{color}
\usepackage[normalem]{ulem}
\usepackage{verbatim}

\newcommand{\ul}{\underline}

\newcommand{\shapename}{\textsf}

\newcommand{\cut}[1]{}

\title{Programmable control of nucleation for algorithmic
  self-assembly \thanks{A preliminary version of this paper appears in~\cite{SchulmanWinfree2004}.
    This work was supported by NSF CAREER Grant No. 0093486 to EW,
    NASA grant NNG06GA50G to EW and NSF Graduate Fellowship to RS.}}

\author{Rebecca Schulman\thanks{Department of Physics, University of California Berkeley, Berkeley, CA, USA} \and Erik Winfree\thanks{Department of Computer Science and Department of Computation and Neural Systems, California Institute of Technology, Pasadena, CA, USA}}

\begin{document} 
\maketitle

\begin{abstract} 
  Algorithmic self-assembly, a generalization of crystal growth
  processes, has been proposed as a mechanism for autonomous DNA
  computation and for bottom-up fabrication of complex nanostructures.
  A `program' for growing a desired structure consists of a set of
  molecular `tiles' designed to have specific binding interactions.  A
  key challenge to making algorithmic self-assembly practical is
  designing tile set programs that make assembly robust to errors that
  occur during initiation and growth.  One method for the controlled
  initiation of assembly, often seen in biology, is the use of a seed
  or catalyst molecule that reduces an otherwise
  large kinetic barrier to nucleation.  Here we show how to
  program algorithmic self-assembly similarly, such that seeded
  assembly proceeds quickly but there is an arbitrarily large
  kinetic barrier to unseeded growth.  We demonstrate this
  technique by introducing a family of tile sets for which we
  rigorously prove that, under the right physical conditions, linearly
  increasing the size of the tile set exponentially reduces the rate
  of spurious nucleation.  Simulations of these `zig-zag' tile sets
  suggest that under plausible experimental conditions, it is possible
  to grow large seeded crystals in just a few hours such that
  less than 1 percent of crystals are spuriously nucleated.
  Simulation results also suggest that zig-zag tile sets could be used
  for detection of single DNA strands.  Together with prior work
  showing that tile sets can be made robust to errors during properly
  initiated growth, this work demonstrates that growth of objects via
  algorithmic self-assembly can proceed both
  efficiently and with an arbitrarily low error rate, even in a model
  where local growth rules are probabilistic.
\end{abstract}

\begin{keywords}
algorithmic self-assembly | DNA nanotechnology | nucleation theory
\end{keywords}
\section{Introduction}

Molecular self-assembly is an emerging low-cost alternative to
lithography for the creation of materials and devices with
sub-nanometer precision~\cite{Whitesidesetal1991,Lehn1993}.  Whereas
top-down methods such as photolithography impose order externally
(e.g., a mask with a blueprint of the desired structure) bottom-up
fabrication by self-assembly requires that this information be
embedded within the chemical processes themselves.

Biology demonstrates that self-assembly can be used to create complex
objects.  Organisms produce sophisticated and
functional organization from the nanometer scale to the meter scale
and beyond.  Structures such as virus capsids, bacterial flagella,
actin networks and microtubules can assemble from their purified
components, even without external direction from enzymes or
metabolism.  This suggests that spontaneous molecular self-assembly
can be engineered to create an interesting class of complex
supramolecular structures.  A central challenge is how to create a
large structure without having to design a large number of unique
molecular components.

Algorithmic self-assembly has been proposed as a general method for
engineering such structures~\cite{Winfree1995} by making use of local
binding affinities to direct the placement of molecules during growth.
The binding of a particular molecule at a particular site is viewed as
a computational or information transfer step.  By designing only a
modest number of molecular species, which constitute the instructions
or program for how to grow an object, complex objects can be
constructed in
principle~\cite{RothemundWinfree2000,SoloveichikWinfree2007,CookRothemundWinfree2003}.
Because a self-assembly reaction occurring in a well-mixed vessel is
inherently parallel, it is necessary to ensure that the molecules that
encode the instructions for assembly execute these reactions in the
correct order.  The primary concern of this paper is how to design a
set of molecules that correctly initiate the execution of a
self-assembly program.  We address this question theoretically, using
a model that is commonly used to study
crystallization~\cite{LeviKotrla1997}, but which incorporates the
particularities of algorithmic self-assembly.

\begin{figure}
\centering
\epsfig{file=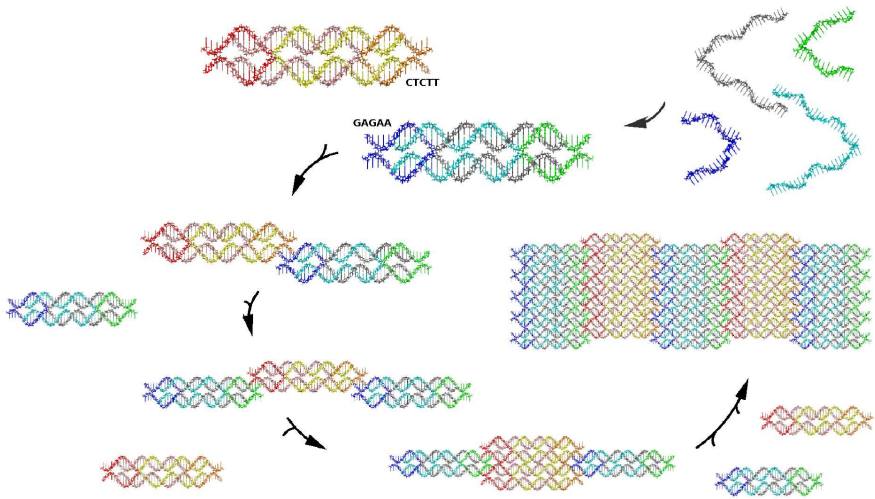,width=12cm}
\caption{{\bf Assembly of DNA strands into DNA tiles and DNA crystal
    lattices.} The configurations are depicted using the NAMOT
  modeling program~\cite{TungCarter1994}.  Stages of an assembly
  reaction during an anneal are separated by successive arrows.
  Strands with
    different sequences are shown in different colors.  At high
  temperatures (first stage) strands are free.  As
    the temperature is lowered, strands assemble into tiles (second
  stage). Each tile displays four sticky
    ends. Example sequences are shown for a pair of complementary
    sticky ends, one on each tile.  As the
    temperature is lowered further, tiles successively join to form
  lattices (third through sixth stages).}
\label{lattice_figure}
\end{figure}

To motivate the model we use, we first describe a specific molecular
system that can implement algorithmic self-assembly experimentally.
DNA double crossover molecules~\cite{Fu93} and related
complexes~\cite{LaBean99a,Mao99b,Yan_etal2003,Heetal2005} (henceforth,
`DNA tiles') have the necessary regular structure and programmable
affinity to implement algorithmic self-assembly, and simple
periodic~\cite{Winfree98c,LaBean99a,SchulmanWinfree2007} and
algorithmic~\cite{Mao2000,Rothemund2004sier,BarishRothemundWinfree2005,Barishetal2009}
self-assembly reactions have been realized experimentally.  As an
example, consider one of the
  DNA double crossover molecules shown in
Figure~\ref{lattice_figure}, which self-assembles from 4 strands of
synthetic DNA.  The  sequences have been designed
such that the desired pseudoknotted configuration maximizes the
Watson-Crick complementarity.  Since the energy landscape for folding
is dominated by logical complementarity more so than by specific
sequence details, it is possible to design similar double crossover
molecules with completely dissimilar sequences.  To date, nearly 100
different molecules of this type have been synthesized.

Interactions between DNA tiles are dictated by the base sequences of
each of four single-stranded overhangs, termed `sticky ends,' which
can be chosen as desired for each tile type.  Tiles assemble through
the hybridization of complementary sticky ends.  The free energy of
association for two tiles in a particular orientation is assumed to be
dominated by the energy of hybridization between their adjacent sticky
ends.  The hybridization energy is favorable when complementary sticky
ends bind, but negligible or unfavorable for non-complementary sticky
ends.  The DNA tiles shown assemble
  (Figure~\ref{lattice_figure}) via the binding of sticky ends to
four adjacent molecules; repeated binding between DNA tiles and
assemblies can produce a lattice.  When multiple tile types are
present in solution, each site on the growth front of the crystal
preferentially will select from solution a tile that makes the most
favorable bonds.  Under appropriate physical conditions, a tile that
can attach by two sticky ends will be secured in place, while tiles
that attach by only a single sticky end usually will be rejected due
to a fast dissociation reaction.  We call these ``favorable'' and
``unfavorable'' attachments, respectively.

\begin{figure}
  \centering
  \epsfig{file=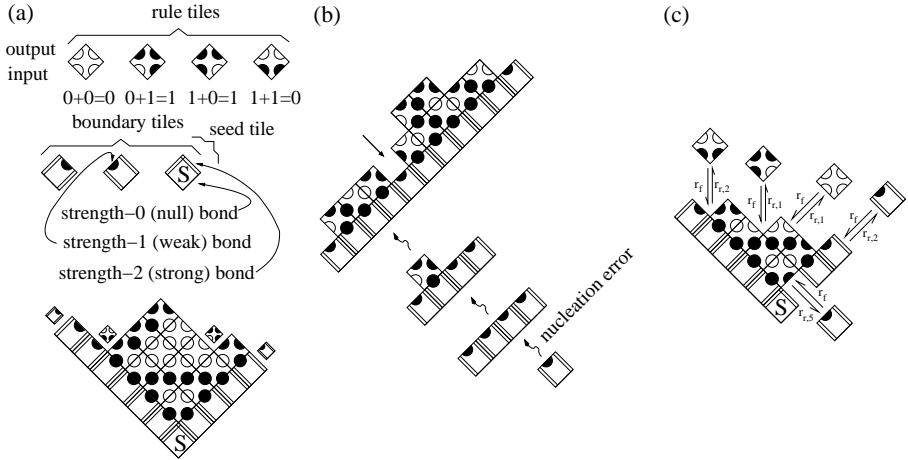,width=12cm}
  \caption{{\bf The Sierpinski tile set.}  {\bf (a)} Because DNA tiles
    are generally not rotationally symmetric, formal tiles cannot be
    rotated. The lower diagram shows the seeded growth of the
    Sierpinski tiles according to the aTAM at $\tau=2$.  The small
    tiles indicate the (only) four sites where growth can occur.  When
    growth begins from a seed, no more than one tile type can attach
    at each location, so assembly always produces the same pattern.
    {\bf (b)} Errors can result from improper
    nucleation when assembly  does not begin from the
    seed tile.  Tile sets containing a tile that can polymerize due to strong
      bonds are particularly prone to nucleation errors.  Improper
    nucleation can produce a long facet where a single insufficient
    attachment can allow a surrounding block of tiles to attach
    favorably.  Different such blocks of tiles may be incompatible,
    leading to an inevitable mismatch at their
    interface.  The straight arrow indicates a
    site where such a mismatch must occur.  {\bf (c)} The rates of
    tile assembly and disassembly in the kinetic Tile Assembly Model
    (kTAM).  For the growth of an isolated crystal under unchanging
    tile concentrations, the forward (association) rate in the kTAM is
    $r_f = k_f [tile] = k_f e^{-G_{mc}}$, while the reverse
    (dissociation) rate is $r_{r,b} = k_f e^{-b G_{se}}$ for a tile
    that makes bonds with total strength $b$.  Parameters $G_{mc}$ and
    $G_{se}$ govern {\ul m}onomer tile {\ul c}oncentration and {\ul
      s}ticky-{\ul e}nd bond strength respectively.  A representative
    selection of possible events is shown here.  Attachments with
    reverse rates $r_{r,1}$ are unfavorable for $G_{mc} > G_{se}$.
    The kTAM approximates the aTAM with threshold $\tau$ when $G_{mc}
    = \tau G_{se} - \epsilon$, for small $\epsilon$.  The same
    set of reactions are favorable or unfavorable in the two models.}
\label{sierpinski}
\end{figure}

The design of an algorithmic self-assembly reaction begins with the
creation of a tile program and its evaluation in an idealized model of
tile interaction, the abstract tile assembly model
(aTAM)~\cite{Winfree1998b}.  A DNA tile is represented as a square
tile with labels on each side representing the four sticky
ends.  Polyomino tiles with labels on each unit-length of the
perimeter can be used in addition to square tiles, since it is
possible to generate the corresponding DNA structures.  A tile program
consists of a set of such tiles, the strength with which each possible
pair of labels binds, a designated seed tile, and a strength threshold
$\tau$.  Under the aTAM, growth starts with a designated assembly of
tiles (usually just the seed tile) and proceeds by allowing favorable
attachments of tiles to occur.  That is, tiles may be added where the
total strength of the connections between the tile and the assembly is
greater than or equal to the threshold $\tau$.  Addition of tiles is
irreversible. At a given step, any allowed
attachment may be performed.  An example of a structure that can be
constructed using algorithmic self-assembly, a Sierpinski triangle, is
shown in Figure~\ref{sierpinski}a.  Beginning with the
seed tile, assembly in the aTAM will result in the growth of an
\shapename{V}-shaped boundary that is subsequently (and
simultaneously) filled in by ``rule tiles''
that obtain their inputs from their bottom sides  and present their outputs on their top
sides.  The four rule tiles for this self-assembly program
have inputs and outputs corresponding to the
four cases in the look-up table for XOR.
  The assembly of these tiles therefore executes the standard
iterative procedure for building Pascal's triangle mod 2.  While
  the Sierpinski triangle construction is particularly simple,
  algorithmic construction is widely applicable: Tile sets for the
construction of a variety of desired products have been
described~\cite{Winfree1995,Lagoudakis1999,Rothemund2000,Adleman2001,CookRothemundWinfree2003,Aggarwaletal2004},
including a tile set capable of universal
construction~\cite{SoloveichikWinfree2007}.

The aTAM captures the essential algorithmic mechanisms of
  generalized crystal growth and makes it possible to program
  self-assembly processes in a straightforward way.  In contrast to
assembly in the aTAM however, the assembly of DNA tiles is
neither errorless nor irreversible, nor is it guaranteed to start from
a seed tile.  For example, in experimental demonstrations of
algorithmic
self-assembly~\cite{Rothemund2004sier,BarishRothemundWinfree2005},
between 1\% and 10\% of tiles mismatched their neighbors and only a
small fraction of the observed crystals were properly nucleated from
seed molecules.  Following~\cite{Schulman_etal2003},
Figure~\ref{sierpinski}b illustrates how unseeded nucleation and
unfavorable attachments can lead to undesired assemblies.

To theoretically study the rates at which errors occur, we need a
model that includes energetically unfavorable events.  The kinetic
tile assembly model (kTAM)~\cite{Winfree1998b} describes the dynamics
of assembly according to an inclusive set of reversible chemical
reactions: a tile can attach to an assembly anywhere that it makes
even a weak bond, and any tile can dissociate from the assembly at a
rate dependent on the total strength with which it adheres to the
assembly (see Figure~\ref{sierpinski}c).  The kTAM is a lattice-based
model, in which free tiles are assumed to be well mixed in solution
and effects within the crystal such as bending or pressure differences
are ignored.  The kTAM has been used to study the trade-off between
crystal growth rate and the frequency of mismatches (errors) in seeded
assemblies~\cite{Winfree1998b,FujibayashiMurata2009}.  One result of these
studies is that, in principle, the rate of mismatch errors can be
reduced by assembling crystals more slowly. Analysis of assembly
within the kTAM also suggests that it is possible to control assembly
errors by reprogramming an existing tile set so as to introduce
redundancy.  ``Proofreading tile
sets''~\cite{WinfreeBekbolatov2003,Chen_Goel2004,Reif_Sahu_Yin2004,SoloveichikWinfree2006}
transform a tile set by replacing each individual tile with a $k\times
k$ block of tiles, exponentially reducing seeded growth errors with
respect to the size of the block.  These results support the notion
that the aTAM, despite its simplicity, provides a suitable framework
for the design of algorithmic crystal growth behavior, {\it ie} that
any tile program for the aTAM can be systematically modified to work
with arbitrarily low error rates in the more realistic kTAM.  However,
previous work did not adequately address the issue of nucleation
errors, which requires extending the kTAM from treating only seeded
growth to treating all reactions occurring in solution.

What is needed is a method of
transforming a tile set to reduce
  the rate of nucleation errors without significant slow-down.
 The transformed tile set must satisfy
two conflicting constraints: when assembly begins from a seed tile, it
must proceed quickly and correctly, whereas assembly that starts
from a non-seed tile must overcome a substantial barrier to
nucleation in order to continue.


How is it possible to have a barrier to nucleation only when no seed
is present?  In a mechanism for the control of 1-dimensional
polymerization, found both in
biology~\cite{SeptMcCammon2001,Collinsetal2004} and
engineering~\cite{DirksPierce2004}, a seed induces a conformational or
chemical change to monomers, without which monomers cannot polymerize.
For example, in spontaneous actin polymerization, it is proposed that
a trimer occasionally bends to form an incipient helix that allows for
further growth~\cite{SeptMcCammon2001}. The Arp 2/3 protein complex
imitates the shape of an unfavorable intermediate of the spontaneous
actin nucleation process~\cite{Kelleheretal1995}.  In contrast,
  in two- and three-dimensional systems---condensation of a gas~\cite{McDonald1962},
crystallization~\cite{Markov2003}, or in  the Ising
model~\cite{Schneidmanetal1999}---classical nucleation
theory~\cite{Zettlemoyer1969,DaveyGarside2000} predicts that a barrier
to nucleation exists because clusters have unfavorable energies
proportional to the surface area of the cluster (possibly due to
interfacial tension or pressure differences with respect to the
surrounding solution), and favorable energies proportional to the
volume of the cluster.  Because volume grows more quickly than surface
area as clusters grow larger, a supersaturated regime exists where
small clusters tend to melt, but above a critical size, cluster growth
rather than melting is favored.  In some crystalline ribbons or tubes,
growth is initially in two dimensions and is
disfavored because of unfavorable surface area/volume interactions, up
to the point that the full width ribbon or tube has been formed.  For
these materials, a seed structure could allow immediate growth by
providing a stable analogue to a full-width assembly.  Protein
microtubules~\cite{Moritzetal1995} and DNA
tubes~\cite{Mitchelletal2004,Rothemund2004a,Liuetal2004} are believed
to exhibit this type of nucleation barrier.

In this paper we describe a tile set family, the zig-zag tile sets,
for the control of nucleation during algorithmic self-assembly.
Zig-zag tiles can assemble immediately on a seed tile to grow
potentially long ribbons of predefined width.  In the absence of a
seed tile, only full-width ribbons can continue to grow exclusively by
favorable attachments.  That is, there is a critical size barrier
(based on unfavorable surface/volume energy interactions) that
prevents spurious nucleation.  By redesigning the tile set it is
possible to increase the width and therefore the critical size. We
prove that in principle this method exponentially reduces the rate at
which assemblies without a seed tile grow large (unseeded growth),
while maintaining the rate of growth that starts from a seed tile an
proceeds roughly according to the aTAM (seeded growth).

Used as part of an error-reducing tile set transformation,
  the zig-zag tiles solve the aforementioned problem
of controlling nucleation during algorithmic growth.
With an appropriate seed, zig-zag ribbons can play the same role as
the \shapename{V}-shaped boundary in Figure~\ref{sierpinski}a.  Since
rule tiles are not likely to spuriously nucleate on their own under
optimal assembly conditions~\cite{Winfree1998b}, once this boundary
has set up the correct initial information, algorithmic self-assembly
will proceed with few spurious side products.

In Section~\ref{zig_zag_tileset_intro}, we describe the zig-zag tile
set family in detail. In Section~\ref{self_assembly_models}, we
introduce a variant of the kTAM that is appropriate for
  the study of nucleation.  In
Section~\ref{zig_zag_growth_and_nucleation} we analyze thermodynamic
constraints on ribbon growth in our model.  In
Section~\ref{asymptotic_analysis}, we prove our main theorem, that the
rate of spurious nucleation decreases exponentially with the width of
the zig-zag tile set.  In contrast, the speed of seeded assembly
decreases only linearly with width.  Thus, for a given volume
we can construct a tile set such that no spurious nucleation is
expected to occur during assembly.  This illustrates how the logical
redesign of molecules can be qualitatively more effective in
preventing undesired nucleation than just controlling physical
quantities such as temperature and monomer concentration.  In
Section~\ref{simulation_section}, we use simulations to provide numerical estimates of nucleation
rates.  These estimates suggest that reasonably
sized zig-zag tile sets can be expected to be effective in the
laboratory.

\section{The Zig-Zag Tile Set}
\label{zig_zag_tileset_intro} 

\begin{figure}[t]
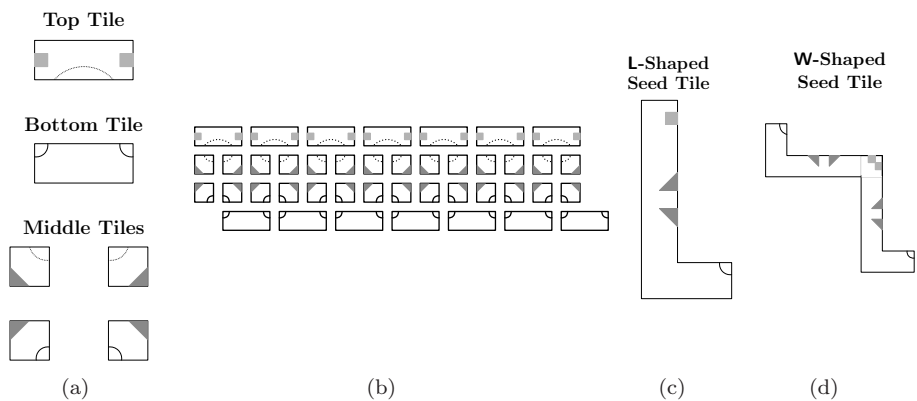

\centering  
\subfigure[\label{tile_program}]{\epsfig{file=dna-nucleation-diagrams.15,width=1.9cm}}
\subfigure[\label{zig_zag_assembly}]{\epsfig{file=dna-nucleation-diagrams.1,width=6cm}}
\subfigure[\label{l_shaped_seed}]{\epsfig{file=dna-nucleation-diagrams.19,width=1.5cm}}
\subfigure[\label{v_shaped_seed}]{\epsfig{file=dna-nucleation-diagrams.11,width=2.3cm}}
\caption{{\bf The zig-zag tile set.}  {\bf (a)} The width 4 zig-zag
  tile set and seed tiles.  Each shape represents a single tile.
  Tiles have matching bonds of strength 1 when the shapes on their
  edges match.  {\bf (b)} The ribbon structure formed by the zig-zag
  tile set.  {\bf (c)} The \shapename{L}-shaped seed nucleates linear
  assemblies.  {\bf (d)} The \shapename{W}-shaped seed tile, with
  appropriate tiles for vertical zig-zag growth, could nucleate
  \shapename{V}-shaped assemblies.}
\label{tiles_figure}
\end{figure}

\begin{figure}[t]
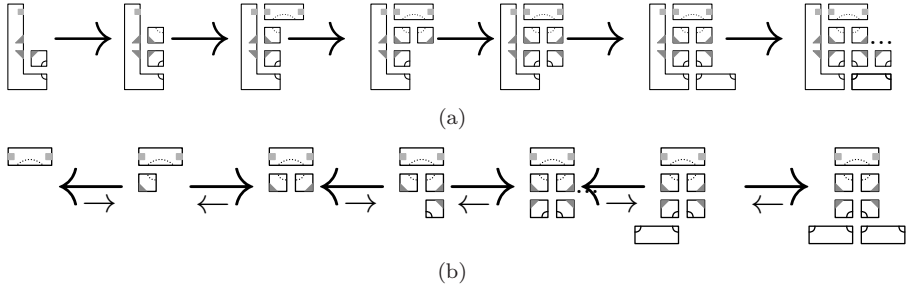

  \centering
  \subfigure[\label{seeded_growth}]{\epsfig{file=dna-nucleation-diagrams.16,width=12cm}}
  \subfigure[\label{unseeded_growth}]{\epsfig{file=dna-nucleation-diagrams.17,width=12cm}}
  \caption{{\bf Zig-zag tile set growth.}  {\bf (a)} Seeded growth of
    a zig-zag tile set in the aTAM.  The same growth pattern occurs
    reversibly in the kTAM when $G_{mc} = 2G_{se} - \epsilon$. {\bf
      (b)} Unseeded growth.  A possible series of steps by which the
    tiles could spuriously nucleate in the kTAM.}
\label{types_of_growth_figure}
\end{figure}

A self-assembly program is a set of tiles that assembles into a
desired shape or set of shapes.  The {\bf zig-zag tile set}
(Figure~\ref{tiles_figure}a of width $k$ contains tiles that assemble
into a periodic ribbon of width $k$ (Figure~\ref{tiles_figure}b).
Zig-zag tile sets of widths $k \ge 2$ can be constructed.  A zig-zag
tile set includes a {\bf top tile} and a {\bf bottom tile}, each
having the same shape as 2 horizontally connected square tiles.  Each
of the $k-2$ rows between the top and bottom tiles contains two unique
{\bf middle tiles} that alternate horizontally.  The alternation of
two tile types along the columns enforces the column-wise staggering
of the top and bottom tiles.  Each tile label has exactly one match on
another tile type, so the tiles cannot assemble to form any other
structures held together by sticky end bonds.

The tile set is designed to operate in a physical regime where the
attachment of a tile to another tile or assembly by two matching sides
is energetically favorable, but an attachment by only one bond is
energetically unfavorable.  In the aTAM, these conditions translate to
growth with a threshold of 2.  Growth beginning from any non-seed tile
in the zig-zag tile set goes nowhere in the aTAM---no two tiles can
join by a bond  with a strength of at least
2. In contrast, growth can proceed from an \textsf{L}-shaped or
\textsf{W}-shaped seed tile
(Figures~\ref{tiles_figure}c~and~\ref{tiles_figure}d).
Figure~\ref{types_of_growth_figure}a illustrates the only possible
growth path in the aTAM from the \textsf{L}-shaped seed.  The
staggering of the top and bottom tiles allows growth to continue
indefinitely along a zig-zag path.  Note that the top and bottom tiles
alternately provide the only way to proceed to each successive column.
Assemblies that do not span the full width ($k$ tiles) either cannot bind top tiles or cannot
  bind bottom tiles, and thus cannot grow indefinitely.  Growth from
a seed tile of less than full width would stall.  For example, with a
seed tile of width $k-1$, the top tile could not attach by two bonds
to the assembly.

In the kTAM, seeded growth occurs in the same pattern as in the aTAM.
Unlike in the aTAM, however, there are also series of reactions that
can produce a full width assembly in the absence of a seed tile.  The
formation of such an assembly is called a spurious nucleation error.
An example of such unseeded growth is shown in
Figure~\ref{types_of_growth_figure}b.
Under the conditions of interest, some steps in spurious nucleation
are energetically favorable, but at least $k-1$ must be unfavorable
before the full-width assembly is formed.  
Once the full-width assembly is formed, further growth is
favorable.  Spurious nucleation is a transition from assembly {\it
  melting}, where assemblies are more likely to fall apart than they
are to get larger, to assembly {\it growth}, where each assembly step
is energetically favorable.  Any assembly where melting and growth are
both energetically favorable is called a critical nucleus.

Classical nucleation
theory~\cite{Zettlemoyer1969,DaveyGarside2000} predicts that the rate
of nucleation is limited by the concentration of the most stable
critical nucleus, $[A_c]$.  Intuitively, because more unfavorable
reactions are required to form critical nuclei in a wider zig-zag tile
set, $[A_c]$ should decrease exponentially with $k$.  This argument is
not rigorous, however, because the number of
  critical nuclei for a zig-zag tile set also increases
  with $k$.  The rate of spurious nucleation is proportional to the
sum of the concentrations of all these critical nuclei.  We will show
in the following sections that despite the
  increase in the number of kinds of critical nuclei that can form as
  $k$ increases, under many conditions nucleation rates do decrease
exponentially with $k$.

\section{The Self-Assembly Model} 
\label{self_assembly_models} 

To analyze the process of tile assembly, we formally describe the
mass-action kTAM.  For a given tile set, kTAM describes the set of
possible assemblies, their reactions and the dynamics of these
reactions.  The kTAM has been previously used to analyze complex tile
programs~\cite{RothemundWinfree2000,SoloveichikWinfree2007}, and is a
general framework for understanding algorithmic self-assembly.  Here,
we extend the kTAM to include polyomino tiles.  We also introduce a
variant of the kTAM in which the concentrations of all possible
assemblies are considered.  This is in contrast to the original kTAM,
which tracks only a single, seeded assembly.
Our extension is appropriate for studying nucleation,
where growth can begin from any tile.  Also in contrast to previous
work with the kTAM, which used stochastic chemical kinetics, we
introduce mass-action kinetics below.  Both mass-action and stochastic
kinetics are accepted models of chemical kinetics~\cite{Dill2002}, but
mass-action is more tractable analytically and
the results of both models generally converge when a large populations
of molecules are considered.  In Section~\ref{simulation_section}, we
find that simulations of nucleation using stochastic kinetics are
consistent with the bounds on nucleation rates we prove using the
mass-action model.

A {\bf tile type} \(\mathbf{t}\) consists of a shape and a set of bond
types on each unit edge of the shape.  The shape is either a unit
square or a polyomino, a finite, connected set of unit
squares\footnote{Here connected means that every unit square in the
  polyomino must have at least one side that abuts the side of another
  unit square in the polyomino. That is, the polyomino's component
  squares cannot be merely diagonally touching.}.  The set of possible
{\bf bond types} is referred to as \( \Sigma \).  A set of tile types
is denoted by $\mathbf{T}$.  A tile (as contrasted with a tile type)
is a tuple of a tile type and a location, which is specified by  $L = (x,y)$, where $L$ is the coordinate location of the
leftmost top unit
  square within the polyomino.  The set of tiles (all possible tile
types in all possible locations) is referred to as \(\mathit{T}\).
 A translation of a tile has the same tile type as
the original. Tiles cannot be
rotated. Tiles that abut vertically or horizontally are {\bf bound} if
they have the same labels on the abutting sides.  A set of tiles is
bound if there is a path of bound tiles between
any two tiles in the set.

An assembly \( A \) is an equivalence class with respect to
translation of a non-overlapping, bound finite set of one or more
tiles.  The set of assemblies is denoted by $\mathcal{A}$ and the set
of assemblies consisting of two or more tiles is denoted
$\mathcal{A}_{2+}$.  We will also use the notation for the set of
tiles, $T$ to refer to the assemblies that have only one tile,
i.e. $\mathcal{A} - \mathcal{A}_{2+}$. A set of tiles $\tilde{A}$ is
considered the canonical representation of $A$ if $\tilde{A} \in A$ and

\begin{align*}
&\forall \langle \mathbf{t}, (x_i,y_i) \rangle \in \tilde{A},  x_i \ge 0 \text{ and } y_i \ge 0 \text{ and } \\
&\exists y, \mathbf{t'} \text{ s.t. } \langle \mathbf{t'}, (0,y) \rangle \in \tilde{A}  \text{ and } \\
&\exists x, \mathbf{t''} \text{ s.t. }\langle \mathbf{t''}, (x,0) \rangle \in \tilde{A} 
\end{align*}

That is to say, the canonical representation uses a
coordinate system such that the reference locations of the tiles just
fit into in the upper right quadrant of the plane with no negative
coordinates.  Note that polyomino tiles may extend into the
other three quadrants, so long as the location of the leftmost top
unit of each polyomino is in the first quadrant.  For an assembly $A$,
\begin{align*} 
width(A) = \max_{x} |y_1 - y_2|\;+\;1 \text{, such that }  
&\langle \mathbf{t_1}, L_1 \rangle, \langle \mathbf{t_2}, L_2 \rangle \in A, \\
&(x,y_1) = L_1, (x,y_2) = L_2 .
\end{align*}
This definition measures width with respect to the reference points
for polyomino tiles, ignoring the extent of the other unit squares
within the polyomino.  \(length(A)\), is defined analogously.
Note that the definitions of $length$ and $width$ given here are
designed to maximize the clarity of the analysis that follows, and
may not be appropriate for other analyses of tile assembly.  The
addition relation is defined between an assembly $A \in
\mathcal{A}_{2+}$ and a tile $t$ so that $A + t = B$ if and only if
$\tilde{A}$ and $t$ are bound but non-overlapping, and $\tilde{A} \cup
t$ is a member of equivalence class $B$.  For the attachment of two
tiles to each other, we need to be careful to correctly
  count the number of ways tiles can attach\footnote{This definition
  is crafted to correctly count the number of distinct ways in which
  tiles can attach to each other such that first, the system will
  satisfy detailed balance given the free energies assigned to tiles
  and assemblies at the end of this section, and second, the dynamics
  of tile interaction will be unchanged if tiles are given irrelevant markings---e.g., 
  if a new tile, with the same binding labels
  as an existing tile, is added and the concentration of both new
  and old tiles are half that of the original, then in the new system the total concentration of 
  both tile types will have the same dynamics as the original tile's concentration in the
  original system.  The definition can be examined by
  considering the number of ways in which different tiles can attach
  to each other.  Two tiles of the same type with the same label on
  all four sides can attach in exactly two distinct ways, two tiles of
  different type but with the same label on all four sides can attach
  in exactly eight ways, and two tiles of different
  types for which the left side of the first tile matches the right
  side of the second tile, but such that all other bonds are
  non-matching, can attach in exactly one way.}.   We consider the set of tile types
$\mathbf{T}$ to be listed in some order. The addition relation is
defined between two tiles $t_1 = \langle \mathbf{t_1}, L_1 \rangle$
and $t_2 = \langle \mathbf{t_2},L_2\rangle$ if either $\mathbf{t_1}$
comes before $\mathbf{t_2}$ in the ordering of tiles or $\mathbf{t_1}
= \mathbf{t_2}$ and for $L_1 = (x_1,y_1)$ and $L_2 = (x_2,y_2)$,
either $y_1 < y_2$ or $x_1 < x_2$ and $y_1 = y_2$ and $t_1 \cup t_2 =
\tilde{A}$, for some $A \in \mathcal{A}$.  In this case, $t_1 + t_2 =
A$.

Bound tiles have a {\bf bond} between them.  The
standard free energy, $G^{\circ}$, of an assembly $A$ is defined as \(
G^{\circ}(A) = -b G_{se} \), where $b$ is the number of bonds in the
assembly and \( G_{se} \) (the sticky end energy) is the
unitless free energy of a single bond.

The dynamics in the kTAM consists of a set of reactions in which
  assemblies grow larger or smaller.
  
  In this paper, we consider all possible accretion reactions:
  reactions either between two tiles or between a tile and an
  assembly.  We also assume that the number of available single tiles
  does not change during the course of assembly (i.e. the reaction is
  ``powered'' by some process or circumstance that keeps monomer
    concentrations constant).

Formally, the set of {\bf powered accretion reactions} are

\begin{multline*}
R = 
\left\{ A + t \rightarrow B + t, \; B \rightarrow A \; : \right. \
\left.  A,B \in \mathcal{A}_{2+},\; \mathit{t} \in \mathit{T},\; A + \mathit{t} = B \right\} 
\; \cup \; \\
\left\{ t_1 + t_2 \rightarrow A + t_1 + t_2, \; A \rightarrow \emptyset  \; : 
 \mathit{t_1},\mathit{t_2} \in \mathit{T},\; A \in \mathcal{A}_{2+},\; \mathit{t_1} + \mathit{t_2} = A\right\}
\end{multline*}

The appearance of single tiles on both sides of the association
reactions and neither side of the dissociation reactions reflects the
powered model's assumptions that the number
of single tiles remains constant.

In the mass-action kTAM, the dynamics of an assembly process are
governed by mass-action kinetics.  Mass-action kinetics is based on an
ideal situation where tiles and assemblies exist in infinite
quantities, and move at random through a solution of infinitely large
volume.  To distinguish the relative abundance of tile types and
assemblies in the system, we use the notion of {\bf concentration},
which denotes the number
of copies of the relevant tile type or assembly within a unit volume.
The concentration of species $A$ is denoted $[A]$.

Tiles and assemblies form or are consumed because of reactions that happen
spontaneously or as a result of collisions between
the reactants.  This leads to the
concentrations of assemblies changing over time. In
a physical reaction vessel, an association reaction (a reaction where
multiple species interact) occurs at a rate proportional to the
frequency with which all the species involved come into physical
contact.  When the possible reactants are well-mixed and moving
randomly through solution, the frequency with which such contact
occurs is proportional to the product of the concentrations of all the
reactant species.  Likewise, dissociation reactions, which have only
one reactant, occur randomly with constant probability per time unit
per molecule. Even though individual reactions occur
stochastically, when the number of particles is infinite, the total
reaction rate is deterministic.

These observations lead to mass-action kinetics, which is an
idealized model of chemical reactions in a
well-mixed vessel~\cite{Dill2002}. The proportionality constant that
relates the product of the concentrations of the reactant species to
the rate at which the reaction occurs is a {\bf rate constant}.  In
general, for a chemical reaction $\sum_{i}n_i S_i \rightarrow \sum_{j}
m_j S_j$ with rate constant $k$, where $S_i$ are chemical species and
$n_i, m_j \in \mathds{Z}^{\ge 0}$ are the reactant and product
stoichiometries (the number of times the reactant or product species
occurs), mass-action dynamics~\cite{Dill2002} predict \(
\frac{d[S_j]}{ds} = k (m_j-n_j)\prod_{i}[S_i]^{n_i}\).  Mass action
reactions occur in parallel, so that dynamics add linearly for
multiple reactions.

In the kTAM, each reaction has a forward rate constant \(k_f\) that we
assume to be the same for all reactions, and a backward rate constant
\(k_r = k_f e^{-\Delta G^{\circ}} \), where \( \Delta G^{\circ} \) is
the difference between the sum of the unitless standard free
energies of the reactants and that of the products (where the standard
free energy of a single tile is 0).  The concentration of all tile
types is held at $e^{-G_{mc}}$. (Identical concentrations are
considered for convenience only; Appendix \ref{detailed_balance_proof}
shows how our formalism can be extended trivially to treat reactions
where species have different concentrations.) Assemblies consisting of
more than a single tile have an initial concentration of 0.  Thus, for
an assembly $A$ at time point $s$,
\begin{multline}  
\frac{d[A]}{ds} =k_f \biggl(\sum_{\substack{A + \mathit{t} \rightarrow
B + \mathit{t},\\
B \rightarrow A \; \in R}} e^{G^{\circ}(B) - G^{\circ}(A)}[B] -
[A]e^{-G_{mc}} \;+ \\
 \sum_{\substack{B + \mathit{t} \rightarrow A + \mathit{t},\\ 
A \rightarrow B \; \in R}} [B]e^{-G_{mc}} -
e^{G^{\circ}(A) - G^{\circ}(B)}[A] \;+ \sum_{\substack{\mathit{t_1} +
\mathit{t_2} \rightarrow A+\mathit{t_1}+\mathit{t_2},\\
A \rightarrow \emptyset \in
R}}e^{-2G_{mc}} -
e^{G^{\circ}(A)}[A] \biggr). 
\label{dA_ds_equation}
\end{multline}
Each term in the first summation is the difference between the rate at
which $A$ and a tile react to form a larger assembly $B$ and the rate
at which the larger assembly $B$ decomposes into $A$ and a tile.  Each
term in the second summation is the difference between the rate of
formation of $A$ by a reaction where a single tile binds to a smaller
assembly $B$, and the rate decomposition of $A$ into assembly $B$ and
a single tile.  The terms in the final summation are the rate of
formation of $A$ from two single tiles and its dissociation into
  two single tiles.  These final terms are nonzero only if $A$ is an
  assembly composed of exactly two tiles.  In the remainder of this
paper, we refer to the mass-action kTAM with powered accretion
reactions as simply ``the kTAM.''

The free energy $G(A)$, (in contrast to the standard free energy
$G^{\circ}(A)$, reflects both the entropy loss due to crystal
formation and the enthalpy gain of assembly.  For an assembly $A$ with
$n$ tiles and $b$ bonds, it is defined as \( G(A) = G^{\circ}(A) + n
G_{mc} \).  The steady state concentration of an assembly $A$ is given
by \( [A]_{ss} = e^{-G(A)} = e^{\left(b G_{se} - n G_{mc}\right)}
\). Recall that $G_{mc} > 0$ and $G_{se} > 0$, so that the
  energetic penalty of adding an additional tile can be compensated
  for by forming sufficiently many new bonds.  A smaller
$G(A)$ is more favorable, and corresponds to a higher steady state
concentration.

  This model satisfies detailed balance within $\mathcal{A}_{2+}$.  That is, for all reaction
  pairs $A \rightarrow B$, and $A + t \rightarrow B + t$, \(
  k_f[t][A]_{ss} = k_r[B]_{ss} \), where $k_f$ and $k_r$ are the
  forward and reverse rates in the respective reactions, and for
  reaction pairs $t_1 + t_2 \rightarrow A$ and $A \rightarrow
  \emptyset$, $k_f[t_1][t_2] = k_r[A]_{ss}$.  A proof that the kTAM satisfies detailed balance is
    contained in Appendix \ref{detailed_balance_proof}.

\section{Thermodynamics of Zig-Zag Assemblies}
\label{zig_zag_growth_and_nucleation}
\begin{figure*}[t]
  \centering
  \subfigure[\label{4_wide_phase_diagram}]{\epsfig{file=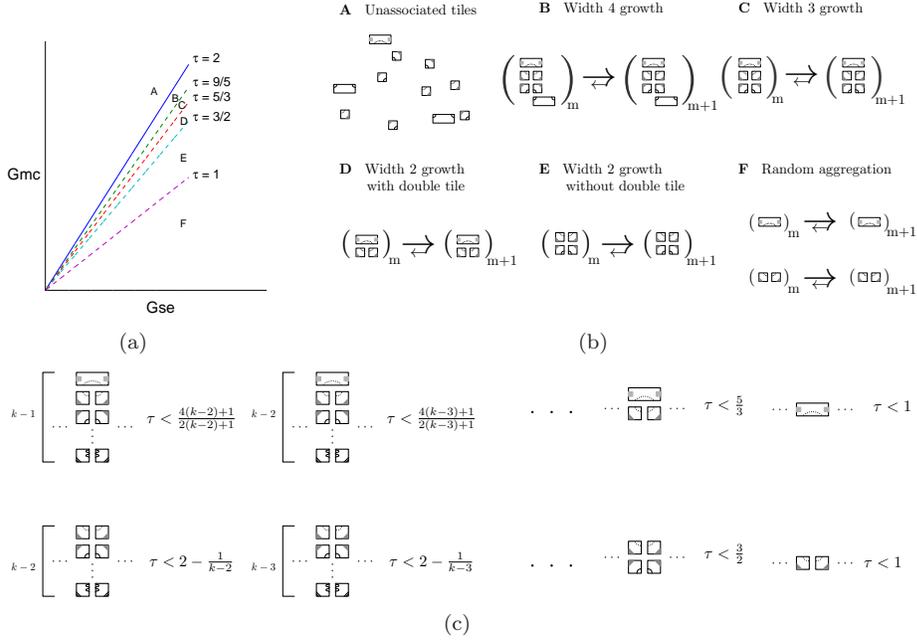,width=3.5cm}}
  \subfigure[\label{4_wide_phase_reactions}]{\epsfig{file=dna-nucleation-diagrams.20,width=8.5cm}}
  \subfigure[\label{general_phase_separations}]{\epsfig{file=dna-nucleation-diagrams.13,width=12cm}}\caption{{\bf
      Physical conditions where zig-zag polymer elongation is
      favorable.}  $G_{mc}$ (ln(tile concentration)) and $G_{se}$
    (bond strength) define a set of physical conditions for zig-zag
    tile assembly. $\tau = \frac{G_{mc}}{G_{se}}$.  {\bf (a)} Phase
    diagram of the width 4 zig-zag tile set.  In phase {\bf A}, above
    the line $\tau=2$, no assembly reactions are favorable, whereas in
    regimes {\bf B},{\bf C},{\bf D},{\bf E} and {\bf F}, progressively
    more types of assemblies (shown in (b)) become favorable.  {\bf
      (b)} The polymeric assemblies which become favorable in the
    regimes {\bf B}-{\bf F} shown in (a).   Polymers shown for earlier
    regimes are also favorable in later phases: the polymer shown for
    regime {\bf B} is favorable in regimes {\bf C}-{\bf F} and so on.
    {\bf (c)} The assemblies that can form from a zig-zag tile set of
    width $k$ and the physical conditions (in terms of $\tau$) in
    which these assemblies becomes favorable.  }
\label{phase_portrait}
\end{figure*}

To prove that nucleation rates of zig-zag ribbons decrease
exponentially as their widths increase, we would first like to
identify the critical nuclei for spurious nucleation.  Thermodynamic
constraints provide a powerful tool: Because undesirable assemblies have unfavorable energies,
we can conclude that they occur rarely without having to consider
rates.  (In contrast, assemblies with favorable energies may or may
not form quickly, depending upon details of the kinetics; such
analyses form the bulk of Sections~\ref{asymptotic_analysis} and
\ref{simulation_section}.)

We therefore consider the free energy landscape, where each point in
the landscape corresponds to a particular type of assembly.  Optimal
control over nucleation is achieved in a regime where zig-zag growth
is  favorable, but the growth of less than full-width
(thin) assemblies is  unfavorable.  

Within the kTAM, the energy landscape for assemblies is formally
described by the free energy $G(A) = n G_{mc} - b G_{se}$, which can
be evaluated directly for any given assembly $A$.  $G_{se}$ and
$G_{mc}$ describe the physical conditions for assembly.  Changing
$G_{se}$ and $G_{mc}$ can bring the system into two qualitatively
different phases.  In the melted phase, $G(A)$ is bounded below by
$G_{mc}$ for all $A$, meaning that no assembly has a concentration of
more than $e^{-G_{mc}}$ at steady state.  In contrast, in the
crystalline phase, $G(A)$ can continue to decrease without bound (so
$[A]_{ss} = e^{-G(A)}$ can increase without bound) as certain
polymeric assemblies become longer and longer -- that is, adding a
repeat unit to the assembly strictly decreases its free
energy\footnote{In powered models, formal steady state concentrations
  can continually increase.  This may seem nonphysical,
  but it is not problematic; it reflects the fact that providing
  unbounded materials can lead to an unbounded accumulation of
  product, and that longer polymers do not achieve steady state within
  the time during which the powered model is an appropriate model.}.
Within the crystalline phase, there are regimes where the elongation
of different types of polymers are favorable or unfavorable.  To
ensure that thin polymers do not tend to grow, it is enough to show
that for each of these polymer types, longer polymers have a higher
free energy than shorter ones.

\begin{figure}
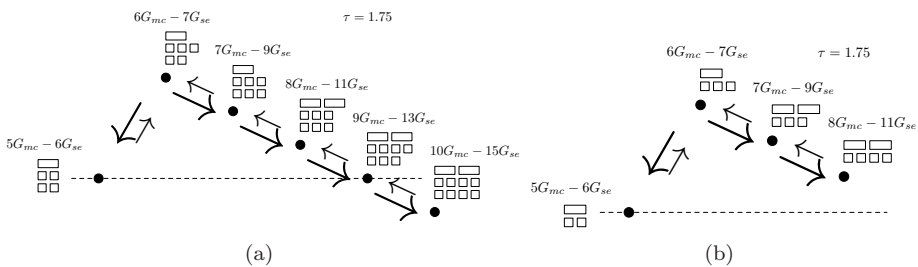

\centering
\subfigure[]{\epsfig{file=dna-nucleation-diagrams.21,width=6.85cm}}
\subfigure[]{\epsfig{file=dna-nucleation-diagrams.22,width=5.15cm}}
\caption{{\bf Zig-zag polymerization reactions.}  The addition of a
  polymer unit to a thin assembly consists of an initial unfavorable
  accretion reaction followed by a series of favorable accretion
  reactions. {\bf (a)} A favorable polymerization reaction.  The
  positive free energy change from the four favorable accretion
  reactions is larger than the negative energy change from the initial
  unfavorable accretion reaction.  Thus, the elongation of polymers of
  width 3 is favorable when $\tau=1.75$.  {\bf (b)}
  An unfavorable polymerization reaction.  The positive free energy
  change from the two favorable accretion reactions is not large
  enough to compensate for the negative energy change from the initial
  unfavorable accretion reaction, so that the elongation of polymers
  of width 2 is unfavorable when $\tau=1.75$. }
\label{lengthening_example_figure}
\end{figure}

To characterize the energy landscape formally, we consider the
important classes of polymeric assemblies and evaluate their free
energies.  Figure~\ref{phase_portrait}b, {\bf B}-{\bf
  F} show the 6 main types of polymeric
assemblies\footnote{``Imperfect'' long assemblies, such as an assembly with more tiles in one column than another, can be
  considered as a member of the  class
  corresponding to a more
  complete assembly of the same length and width.  Since
  removing tiles from a ``perfect'' assembly strictly increases its
  free energy, these ``imperfect'' assemblies have strictly lower
  concentrations than their corresponding ``perfect'' assembly.} for
ribbons of width 4 by indicating the repeat group that may be added
(by a series of accretion reactions, as shown in
Figure~\ref{lengthening_example_figure}) to extend the polymer.  To
determine whether adding a repeat group results in a higher or lower
energy assembly, we evaluate $\Delta G = G(A_{m+1}) - G(A_m) = \Delta
n G_{mc} - \Delta b G_{se}$ where $A_m$ is a polymeric assembly with
$m$ repeat units.  If $\Delta G$ is negative, then longer polymeric
assemblies of this type are more favorable and we can expect this kind
of assembly to grow at some rate.  This gives a linear condition on
$G_{se}$ and $G_{mc}$, specifying a regime of physical conditions in
which a certain class of long assembly is favorable.  For example, for
polymer type {\bf E}, each repeat unit adds 4 tiles ($\Delta n = 4$)
and 6 bonds ($\Delta b = 6$) , so these polymers grow if $ 4 G_{mc} -
6 G_{se} < 0$, i.e.  $\frac{G_{mc}}{G_{se}} < \frac{3}{2}$.  Similar
calculations result in the phase diagram shown in
Figure~\ref{phase_portrait}a, which shows the melted
phase {\bf A}, in which no polymers are favorable, and the crystalline
phase divided into regimes {\bf B}-{\bf F} wherein one additional type
of polymer becomes favorable in each successive regime.  In all these
calculations, the ratio $\tau \stackrel{def}{=} \frac{G_{mc}}{G_{se}}$
plays a critical role.

Figure~\ref{phase_portrait}c shows the $2k-3$ classes of polymeric
assemblies for the width $k$ zig-zag tile set (excluding the full
width ribbons) along with the condition on $\tau$ that determines when
polymer elongation is favorable.  Exclusively the
elongation of full-width ribbons is favorable when $2 > \tau >
\frac{4(k-2)+1}{2(k-2)+1} = 2 - \frac{1}{2k-3}$.  That is, when $2 >
\tau > 2 - \frac{1}{2k-3}$, zig-zag growth is favorable, but the
elongation of all less than full-width polymers is unfavorable.  The
regime where $2 > \tau > 2 - \frac{1}{2k-3}$ will be referred to as {\bf
  optimal nucleation control conditions.}

\begin{figure}
\centering
\epsfig{file=dna-nucleation-diagrams.23,width=8cm}
\caption{{\bf Reactions that increase width, the $\Delta G$ for those
    reactions, and the resulting conditions (in terms of $\tau$) where
    those reactions are favorable.}}
\label{widening_tau_figure}
\end{figure}

The table in Figure~\ref{widening_tau_figure} enumerates the
  assemblies for which growing wider (rather than longer) is
  favorable.   Like the polymerization of thin ribbons, a
  reaction to produce a wider assembly from a thinner one consists of
  an initial unfavorable accretion reaction followed by a series of
  favorable accretion reactions to complete the new row.  The number
  of favorable reactions determines the values of
    $\tau$ for which the overall reaction is favorable.  Very long
but thin assemblies can favorably grow wider even when $\tau$ is
close to 2, so for optimal nucleation control it is necessary that
elongation of thin assemblies be unfavorable.
Otherwise, a favorable path to nucleation exists: an assembly can grow
longer until it is favorable for it to grow wider and then
it can grow to full width.

\begin{figure}
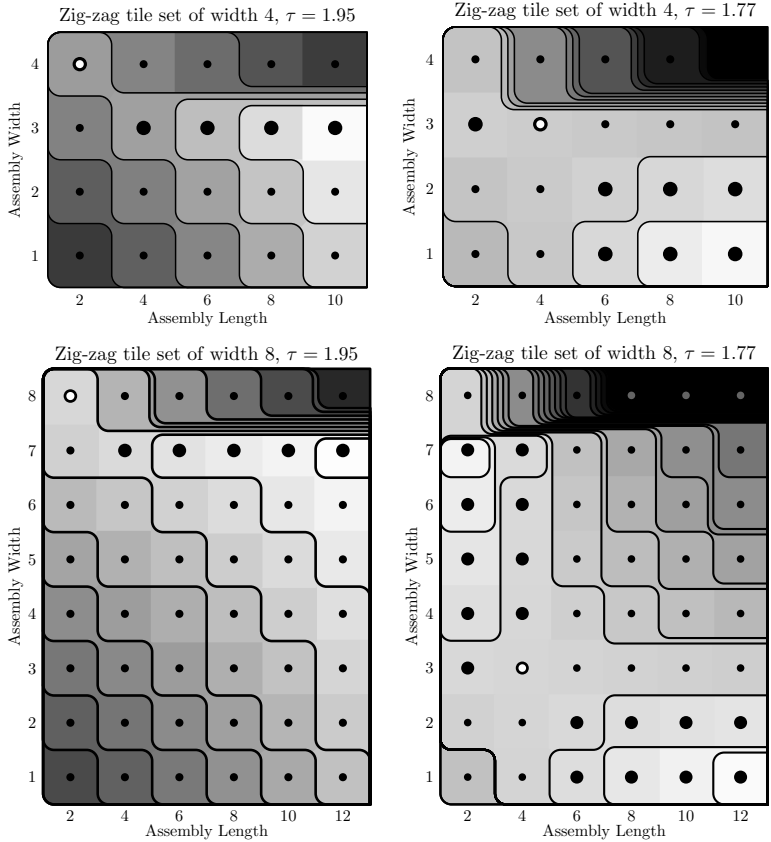

\centering
\epsfig{file=dna-nucleation-diagrams.25,width=5cm}
\epsfig{file=dna-nucleation-diagrams.27,width=5cm}
\epsfig{file=dna-nucleation-diagrams.26,width=5cm}
\epsfig{file=dna-nucleation-diagrams.28,width=5cm}
\caption{{\bf Example energy landscapes.}  Coarse-grained depictions
  of the energy landscapes for two zig-zag tile sets of different
  widths under two different physical conditions.  Each square in the
  grid represents a ``perfect'' assembly of the labeled
  width and length.  The shading in the square corresponding to each
  width and length represents the energy of a rectangular assembly of
  those dimensions.  Darker is more favorable.  Contour lines group
  assemblies of similar energies.  Large circles denote assembly sizes
  that are critical nuclei.  The most favorable critical nucleus (the
  principal critical nucleus) is denoted by a large hollow circle. }
\label{energy_landscapes_figure}
\end{figure}

An example of the difference in the energy landscape between the
regime where only the elongation of full width polymers is favorable
(optimal nucleation control conditions), and a regime where the growth
of thinner polymers is also favorable can be seen in
Figure~\ref{energy_landscapes_figure}.  In each landscape, the {\bf
  critical nuclei} divide the energy landscape into two basins whose
lowest energy assemblies are infinite polymers or fully melted,
respectively.  A critical nucleus can, via a series of energetically
favorable increases or decreases in length or width, either reach full
width or melt away.  The principal critical nucleus is the most stable
critical nucleus.  We start by considering the two landscapes under
optimal nucleation control conditions, the two left landscapes of
Figure~\ref{energy_landscapes_figure}. In these landscapes, the
critical nuclei are of width $k-1$ (or width $k$) for both tile set
widths and the most favorable path to nucleation for both tile sets is
for a crystal of length 2 to grow to full width. Thus, the barrier to
nucleation for a tile set of width 8 is higher than the barrier to
nucleation for a tile set of width 4.  In contrast, when $\tau=1.77$,
the principal critical nucleus is the same for both tile sets: it is
an assembly of width 3 and length 4.  Under these conditions, the
spurious nucleation rate of the tile set of width $k=8$ will not be
appreciably smaller than nucleation rate of the tile set of width
$k=4$.

The primary theorem of the next section will apply only under optimal
nucleation control conditions.  While this region covers only a small
fraction of area in the phase diagram shown in
Figure~\ref{widening_tau_figure}, a slow anneal from a high
temperature where $\tau \gg 2$ to a temperature in which $\tau < 1$
will pass through this regime, and a slow enough anneal will allow the
bulk of the reaction to take place in this regime.  Therefore, it is
reasonable to consider a mechanism for the control of nucleation which
is valid only in this narrow range of physical conditions.  In the
next section, we analyze the nucleation rates of the zig-zag tile set
within this regime.

\section{An Asymptotic Bound on Spurious Nucleation Rates}
\label{asymptotic_analysis}

The kinetic Tile Assembly Model predicts the concentration of each
assembly at all times.  For most tile sets, the number of possible
assemblies is large, and the individual concentrations of many kinds
of intermediate assemblies are not necessarily of
interest. However, the sheer number of possible assemblies and
  possible assembly pathways can significantly affect the overall rate
  of spurious nucleation, and they can not simply be ignored.
  Understanding the contribution of many different assembly types to
  the total spurious nucleation rate is the main technical challenge
  in what follows.  It is often helpful to
talk about the concentration of a {\bf class} $\mathcal{C} \subset
\mathcal{A}$ of assemblies, \( [\mathcal{C}] = \sum_{A \in
  \mathcal{C}} [A] \).  The derivative of the concentration of a class
of assemblies, \( \frac{d[\mathcal{C}]}{ds} = \sum_{A \in \mathcal{C}}
\frac{d[A]}{ds} \), can be calculated as the difference between the
rate at which at which assemblies join the class and that at which
they leave the class.  Reactions which produce new members of the
class from assemblies not in the class are the {\bf inward perimeter
  reactions}, \( R^{in} = \{ A + t \rightarrow B + t,\; A \rightarrow
B,\; t_1 + t_2 \rightarrow B + t_1 + t_2 : A \notin \mathcal{C},\; B
\in \mathcal{C} \} \).  Reactions which use up members of the class to
produce assemblies not in the class (or single tiles) are the {\bf
  outward perimeter reactions} = \( R^{out} = \{ B + t \rightarrow A +
t,\; B \rightarrow A,\; B \rightarrow \emptyset :\; A \notin
\mathcal{C},\; B \in \mathcal{C} \} \).

Define the {\bf flux} across a set of reactions $R$ at time $s$ as 

\begin{multline}
\label{flux_expression}
F(R,s) = \sum_{A + t \rightarrow B + t \in R} k_f [A]_se^{-G_{mc}} +
\sum_{B \rightarrow A \in R} k_f e^{G^{\circ}(B) - G^{\circ}(A)}[B]_s\; + \\
\sum_{t_1 + t_2 \rightarrow A + t_1 + t_2 \in R} k_f e^{-2G_{mc}} + 
\sum_{A \rightarrow \emptyset \in R} k_f e^{G^{\circ}(A)}[A]_s
\end{multline}

where $[A]_s$ is the value of $[A]$ at time point $s$.  Then
\( \frac{d[\mathcal{C}]}{ds}(s) = F(R^{in},s) - F(R^{out},s) \).

We will use these formalisms to bound the rate of spurious nucleation
in a zig-zag tile set of width $k$. The {\bf spuriously nucleated
  assemblies} for a zig-zag tile set of width $k$ will be
denoted $\mathcal{C}_k$.  Let the top tile in Figure
\ref{tile_program} be designated \(\mathbf{t_t}\), the
bottom tile be \(\mathbf{t_b}\), and the seed tile be designated
  \(\mathbf{t_s}\).  Formally,
\begin{multline} 
\mathcal{C}_k = \{ A \in \mathcal{A} : \;\exists (x,y), (w,z) \in
\mathds{Z}^2 
\text{ s.t. } A(x,y) = \mathbf{t_t}, A(w,z) = \mathbf{t_b}, \; \text{ and } \\
\forall (q,r) \in \mathds{Z}^2,\;\; A(q,r) \neq \mathbf{t_s} \}
\end{multline}
Note that the assemblies in $\mathcal{C}_k$ do not contain a seed
tile, and we are measuring the rate of formation of zig-zag ribbons without seed
  tiles.

\begin{figure}[t]
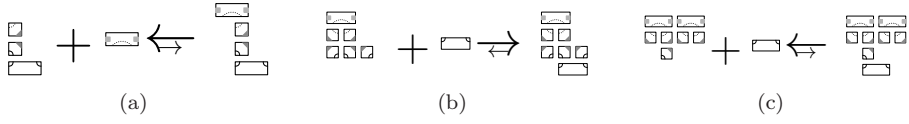
 
\centering 
\subfigure[\label{spurious_reaction_1}]{\epsfig{file=dna-nucleation-diagrams.6,width=3.5cm}}
\hspace{.5cm}
\subfigure[\label{spurious_reaction_2}]{\epsfig{file=dna-nucleation-diagrams.7,width=3.5cm}}
\hspace{.5cm}
\subfigure[\label{spurious_reaction_3}]{\epsfig{file=dna-nucleation-diagrams.8,width=3.5cm}}
\caption{{\bf Spurious nucleation reactions.} Three spurious
  nucleation reactions for a zig-zag tile set of width 4.  
  The reaction  may be either favorable or unfavorable.  In (b), the
  addition is favorable when $G_{mc} = 2G_{se} - \epsilon$ for small $\epsilon$, because two new
  bonds are formed; in (a) and (c), the addition is unfavorable,
  because in each reaction only one new bond is formed.  
}
\label{spurious_nucleation_reactions}
\end{figure}

The inward perimeter reactions for $[\mathcal{C}_k]$, which we call
the {\bf spurious nucleation reactions} and denote by $R^{in}_k$, are
the reactions for which the product is a full width assembly, but the
reactant is not.  In other words, they are the addition reactions
which produce width $k$ assemblies from assemblies of width $k-1$ by
adding either a top or bottom double tile
(Figure~\ref{spurious_nucleation_reactions}).  As shown in
Section~\ref{zig_zag_growth_and_nucleation}, under optimal
  nucleation control conditions these reactions demarcate the point
at which sustained growth can proceed by exclusively favorable steps.
The outward perimeter reactions, which we call the {\bf ribbon
  shrinking reactions} and denote by $R^{out}_k$, are those in which a
tile falls off a full width assembly to produce an assembly of width
$k-1$. For assemblies that have suffered a ribbon shrinking reaction,
there is also a downhill path to complete melting in
an energy landscape of the type shown in
Figure~\ref{energy_landscapes_figure} under optimal
  nucleation control conditions.

The overall rate of spurious nucleation of width $k$ zig-zag crystals
(in units of molar per second),
$$n_k(s) = \frac{d[\mathcal{C}_k]}{ds}(s) = F(R^{in}_k,s) - F(R^{out}_k,s),$$
may be integrated over time to obtain the total concentration of
spuriously nucleated assemblies.  Furthermore, an upper bound on
$n_k(s)$ similarly translates into an upper bound on the concentration
of spuriously nucleated assemblies.  Because the growth path for
full-width ribbons is so favorable (zig-zag growth), one such bound is
obtained by neglecting the ribbon shrinking reactions and
considering just the spurious nucleation reactions:
$$n^+_k(s) = F(R^{in}_k,s) > n_k(s).$$
In what follows, ``the rate of spurious nucleation'' refers to
$n_k(s)$, the rate at which the concentration of 
spuriously nucleated assemblies increases.  We distinguish this rate
from $n^+_k(s)$, the rate of formation of all full-width
assemblies, whether they later shrink or not,  by
referring to the latter as ``the rate of spurious nucleation reactions
(or events).''
\\
\begin{theorem} 
\label{powered_accretion_theorem} 
For a zig-zag tile set of width $k > 2$, if $2 > \frac{G_{mc}}{G_{se}}
> 2 - \delta$, $\delta < \frac{1}{2k-3}$, and \( G_{se} >
\frac{2k\ln{2}}{1-(2k-3)\delta}\), then for all times $s$, \(
n_k(s) < 4k_f e^{(\delta-k) G_{se}} \).
\end{theorem}
\\
\begin{proof}  
  Since \( n_k^+(s) < n_k(s)\), we can prove the theorem by showing
  that \( n_k^+(s) < 4k_f e^{(\delta-k) G_{se}} \). All
  the spurious nucleation reactions are addition reactions, so if we
  compute $n_k^+(s)$ using Equation~\ref{flux_expression}, the second
  and fourth terms of the expression are both zero.  Spuriously
  nucleated assemblies are defined as assemblies of width $k$, so the
  reactants in the spurious nucleation reactions are of width $k-1$
  (only accretion reactions are allowed).  For a tile set of width $k
  > 2$, the third term of Equation \ref{flux_expression}---the
  contribution of the interaction of two tiles---also drops out.
  Therefore, for a tile set of width $k > 2$ with spurious nucleation
  reactions $R^{in}_k$,
\begin{equation} 
n_k(s) \le n^+_k(s)  = \sum_{A + \mathit{t} \rightarrow B + \mathit{t} \in
R^{in}_k} k_f[A]e^{-G_{mc}},
\label{spurious_nucleation_rate}
\end{equation}
where $[A]$ is the concentration of assembly $A$ at time point $s$. 

While it is in general difficult to calculate $[A]$ at
  an arbitrary time point, the following lemma shows that the
concentration of an assembly can be bounded by its concentration at
steady state, which is easy to compute:
\\
\begin{lemma} 
\label{powered_accretion_steady_state_bound_lemma} 
In a mass-action powered accretion kTAM, if in the initial state only single tiles
  have a positive concentration,  then every assembly has a
concentration less than or equal to its steady state concentration
at all time points\footnote{The concentration of the class
    $\mathcal{C}_k$, which at the conditions we consider contains an
    infinite number of assemblies, is actually infinite at steady
    state.  The inward flux, as we will show, is finite because the
    concentration of unnucleated assemblies stays finite at steady
    state, even though there are also an infinite number of
    unnucleated assemblies.}.
\end{lemma}
\\
\begin{proof} See Appendix \ref{steady_state_bound_proof}. \end{proof}\\

Lemma~\ref{powered_accretion_steady_state_bound_lemma} implies that
\begin{equation*} 
F(R^{in}_k,s) \le \sum_{A + \mathit{t} \rightarrow B + \mathit{t}
\in R^{in}_k} k_f[A]_{ss}e^{-G_{mc}} 
\end{equation*}
where $[A]_{ss}$ is the concentration of assembly $A$ at steady state.

Partitioning the summation according to the length of the reactant
assembly gives
\begin{equation}
 F(R^{in}_k,s) \le \sum_{l=1}^{\infty}
 \sum_{\substack{length(A)=l\\
A + t\rightarrow B + \mathit{t}  \in R^{in}_k}} k_f[A]_{ss}e^{-G_{mc}}. 
\label{steady_state_nucleation_rate}
\end{equation}

\begin{figure}[t]
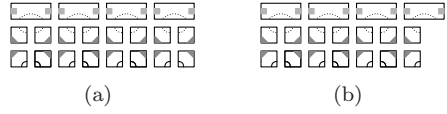
 
\centering \mbox{
\subfigure[\label{even_k_minus_1_by_l}]{\epsfig{file=dna-nucleation-diagrams.4}}
\quad\quad
\subfigure[\label{odd_k_minus_1_by_l}]{\epsfig{file=dna-nucleation-diagrams.5}}
}
\caption {{\bf Assembly dimensions of rectangular assemblies.} {\bf
    (a)} A $k-1=3$ by $l=8$ assembly. {\bf (b)} A $k-1=3$ by $l=7$
  assembly.}
\label{k_minus_1_by_l_rectangles} 
\end{figure} 

To be a reactant in a spurious nucleation reaction, \(A\) must have a
width of $k-1$.  Because all bonds in an zig-zag tile set assembly are
unique to a given location within the ribbon repeat unit, any
potential tile addition either matches the assembly on all sides, such
that no errors occur, or matches on no sides, such that the addition
does not produce a bound assembly. Thus, $A$
cannot have any mismatches.  Each assembly $A$ in the preceding
summation can therefore be viewed as a $k-1$ by $l$ rectangular
assembly of one of the types shown in Figure
\ref{k_minus_1_by_l_rectangles} with zero or more tiles
missing\footnote{It could also be a subset of a rectangular assembly
  with top instead of bottom tiles, but the free energy of both kinds
  of assemblies is the same. To account for this, we include a 2\
  pre-factor in the number of assemblies corresponding to a $k-1$ by
  $l$ rectangle, thereby counting both assemblies with top tiles and
  assemblies with bottom tiles.}. $2G_{se} > G_{mc}$ by assumption, so
the free energy of a $k-1$ by $l$ assembly cannot be more favorable
than the free energy of the $k-1$ by $l$ rectangle that contains it,
since any missing tiles in the rectangle could be added by favorable
reactions.  Therefore, the concentration of any $k-1$ by $l$ assembly
at steady state must be no larger than the concentration of its
corresponding $k-1$ by $l$ rectangular assembly.  Note that this bound
is very loose, since most assembly types have several tiles attached
by only one bond and therefore have a higher free energy.  Let
$A_{k-1,l}$ be a $k-1$ by $l$ rectangular assembly, and $C(k-1,l)$ be
the number of assemblies of width $k-1$ and length $l$.  Each assembly
can bind a single tile in up to $l/2$ locations (since the tile must
be a double tile) along either the top or bottom edge.  Thus,

\begin{align}
 F(R^{in}_k,s) &< \sum_{l=1}^{\infty}
 \sum_{\substack{length(A)=l\\
A + t \rightarrow B + \mathit{t} \in R^{in}_k}}
 k_f[A_{k-1,l}]_{ss}e^{-G_{mc}}\\ 
&\le \sum_{l=1}^{\infty} C(k-1,l) \frac{l}{2} k_f [A_{k-1,l}]_{ss}e^{-G_{mc}}. 
\end{align}

A counting argument shows that \(C(k-1,l) < 2^{(k-1)l + 1}\), so 

\begin{align}
 F(R^{in}_k,s) < \sum_{l=1}^{\infty} 2^{(k-1)l} l k_f
 [A_{k-1,l}]_{ss}e^{-G_{mc}}. 
\end{align}

The steady state concentration of an unseeded assembly with $n$ tiles
and $b$ bonds is given by \( [A]_{ss} = e^{-n G_{mc} + b G_{se}} \).
The assembly $A_{k-1,l}$ contains $(k-2)l$ small tiles and $\lceil l/2
\rceil$ top (or bottom) tiles.  There are $(l-1)(k-2)$ horizontal
bonds between small tiles and $\lceil l/2\rceil-1$ horizontal bonds
between large tiles. In addition, there are up to $l$ vertical bonds
in each of the $k-2$ spaces between rows of tiles.  Therefore,
\begin{equation*}
  [A_{k-1,l}]_{ss} \le \exp\left(-\left((k-2)l + l/2 \right)G_{mc} \; + \right. 
   \left. \left((k-2)(l-1)+l/2+(k-2)l\right)G_{se}\right).
\end{equation*}


Applying the assumption $G_{mc} > (2-\delta)G_{se}$ and simplifying,
\begin{align*}
 [A_{k-1,l}]_{ss} < &\exp\left[(2-k)G_{se} + (k\delta - \frac{1}{2} - \frac{3\delta}{2})l G_{se}\right].
\end{align*}

Thus,
\begin{equation*}
 F(R^{in}_k,s) <\; k_f e^{-G_{mc}} e^{(2-k)G_{se}} 
\sum_{l=1}^{\infty} l 2^{(k-1)l} e^{(k\delta - \frac{1}{2} - \frac{3\delta}{2})l G_{se}}
\end{equation*}

Since $k\delta - \frac{1}{2} - \frac{3\delta}{2} < 0$ when
$k > 2$ and $\delta < \frac{1}{2k-3}$, bounding $G_{se}$ from below preserves
the inequality.  Therefore, when \( G_{se} >
\frac{2k\ln(2)}{1-(2k-3)\delta} \),

\begin{align}
 F(R^{in}_k,s) <&\; k_f e^{-G_{mc}} e^{(2-k)G_{se}} \sum_{l=1}^{\infty} l 2^{(k-1)l} e^{(k\delta - \frac{1}{2} - \frac{3\delta}{2})l\frac{2k\ln(2)}{1-(2k-3)\delta}}\\
=&\; k_f e^{-G_{mc}} e^{(2-k)G_{se}} \sum_{l=1}^{\infty} l 2^{(k-1)l} e^{-kl \ln{2}} \\
=&\; k_f e^{-G_{mc}} e^{(2-k)G_{se}} \sum_{l=1}^{\infty} l 2^{-l} \\
=&\; 2 k_f e^{-G_{mc}} e^{(2-k)G_{se}} \\
\label{proof_before}
=&\; 2 k_f e^{(\delta-k)G_{se}}. \\
\end{align}

\end{proof}
\vspace{4mm} This theorem says that the spurious nucleation rate,
$n_k$, decreases exponentially with $k$ and with $G_{se}$, within the
limits of applicability of the theorem---which requires larger
$G_{mc}$ for larger $k$, and hence slower growth rates.  The strength
of the theorem, therefore, lies in the extent to which spurious
nucleation decreases {\it faster} than the growth rate, $r_k$, of
seeded crystals.  These relative rates translate into the degree of
purity that can be obtained when attempting to grow seeded crystals:
suppose the concentration of seeds is $c$, and they are grown to a
length $L$ during a time period $s = L/r_k$.  The concentration of
unseeded crystals that will have spuriously nucleated in that time is
less than $s \cdot n_k = L \cdot \frac{n_k}{r_k}$, i.e., the
fraction of crystals that were spuriously nucleated is less
  than $\frac{L}{c} \cdot \frac{n_k}{r_k}$.  (When we use $n_k$
without specifying a particular time, we mean its steady state value, which is an upper bound.)
Regardless of what length or amount of seeded crystals is desired,
reducing $\frac{n_k}{r_k}$ is the relevant metric for increasing the
yield of desired structures.

One way to study the trade-off between $n_k$ and $r_k$ is to ask,
given a target growth rate $r$, what is the lowest nucleation rate
that can be achieved by adjusting $G_{mc}$ and $G_{se}$ while
maintaining $r_k = r$?  Previous work \cite{Winfree1998b} has shown
that near the $\tau=2$ phase boundary that is relevant to our theorem,
the growth rate is closely approximated by $$r_k = \frac{k_f}{k-1} (e^{-G_{mc}} -
e^{-2G_{se}})$$ measured in layers per second.  The lowest nucleation
rate for a given target growth rate $r$ is then 
$n^*_k(r) =  \min_{\substack{G_{se}, G_{mc} \\ \hbox{\ s.t.\ } r_k=r}} n_k$. 
A
plot of $n^*_k(r)$ vs $r$, if it could be calculated, would reveal how
much the spurious nucleation rate decreases when the growth rate is decreased.
Theorem~\ref{powered_accretion_theorem} only gives us an upper bound
on $n^*_k(r)$, but even so, this already gives us a characterization
of the advantage provided by wider zig-zag crystals.

Specifically, choosing $2G_{se}-G_{mc}=\epsilon = \ln k$, $\delta = \frac{1}{2} \left(\frac{1}{2k-3}\right)$ and $G_{se} >
4k \ln k$, then
Equation~\ref{proof_before}
guarantees that
\begin{align*}
n^*_k < n_k &< 2 k_f e^{- G_{mc}} e^{(2-k) G_{se}} \\
&= 2 k_f e^{-(G_{mc}-2G_{se})}e^{-k G_{se}} \\
&= 2 k_f e^{\epsilon}e^{-k G_{se}} \\
&= 2 k_f e^{\ln{k}}e^{-k G_{se}} \\
&= 2 k_f k e^{-k G_{se}}. 
\end{align*}
Define $n^B_k = 2 k_f k e^{-k G_{se}}$.  Also
\begin{align*}
r_k = \frac{k_f}{k-1} (e^{-G_{mc}} - e^{-2G_{se}}) = \frac{k_f}{k-1} (e^{\epsilon-2G_{se}} - e^{-2G_{se}}) = k_f e^{-2G_{se}}.
\end{align*}

The ratio $\frac{n_k}{r_k}$ describes the trade-off between
  assembly speed $r_k$ and spurious nucleation $n_k$.  This ratio can
  be no larger than $\frac{n^B_k}{r_k}$.  For $k > 2$ and the chosen
  parameters,
\begin{align*}
\frac{n^B_k}{r_k} = \frac{2k_f k e^{-k G_{se}}}{k_f e^{-2 G_{se}}} = 2k e^{(2-k) G_{se}} < 2k e^{(2-k) 4k \ln{k}}.
\end{align*}
which decreases exponentially with $k$.   Thus, under these
conditions, seeded zig-zag crystals can be grown with exponentially
greater yield as width increases.

\begin{figure}[th]
  \centering
  \epsfig{file=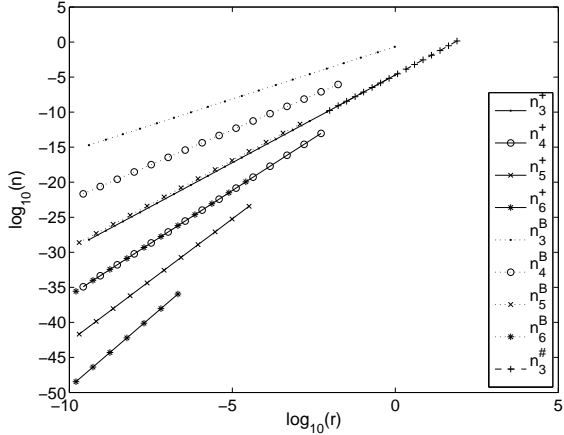,width=7.5cm}
  \caption{{\bf Analytical upper bounds, steady-state calculations and
      asymptotic simulations of spurious nucleation rates.}  The graph
    compares the growth rate $r_k$ (in layers/s) and the rate of
    spurious nucleation events, $n^+_k$ (in M/s), for $2G_{se} -
    G_{mc} = \epsilon = 0.1$. $k_f = 6\times10^5/M/s$ and for all
    points $G_{se} > \frac{2k
        \ln{2}}{1-(2k-3)\delta}$ for $k =3,4,5$ and $6$.  Analytical
    upper bounds on the nucleation rate ($n^B_k =
    4k_fe^{(\delta-k)G_{se}}$) are those given by
    Theorem~\ref{powered_accretion_theorem}.  The method of numerical
    calculation of spurious nucleation reaction rates at steady
    state, $n^+_k$, is described in
    Section~\ref{steady_state_section}.  Stochastic simulation methods
    (giving $n^\#_k$) are described in Section~\ref{instantaneous_nucleation_section}.}
  \label{steady_state_nucleation}
\end{figure}

The bound $n^B_k$ where $\epsilon = 0.1$ is plotted against $r_k$ in
Figure~\ref{steady_state_nucleation} for $k=3,\;4,\;5$ and $6$.  While
these bounds characterizing the trade off between $n^*_k$ and $r_k$
are rigorous, because Theorem~\ref{powered_accretion_theorem} is so
loose, it is expected that $n^*_k$ is actually much lower than the
bound $n^B_k$.  In the following sections, we will
see that this is true; furthermore, the true slopes are even steeper
than obtained by Theorem~\ref{powered_accretion_theorem}.

\section{Numerical Estimates of Spurious Nucleation Rates}
\label{simulation_section}

\begin{figure}[th]
  \centering
  \epsfig{file=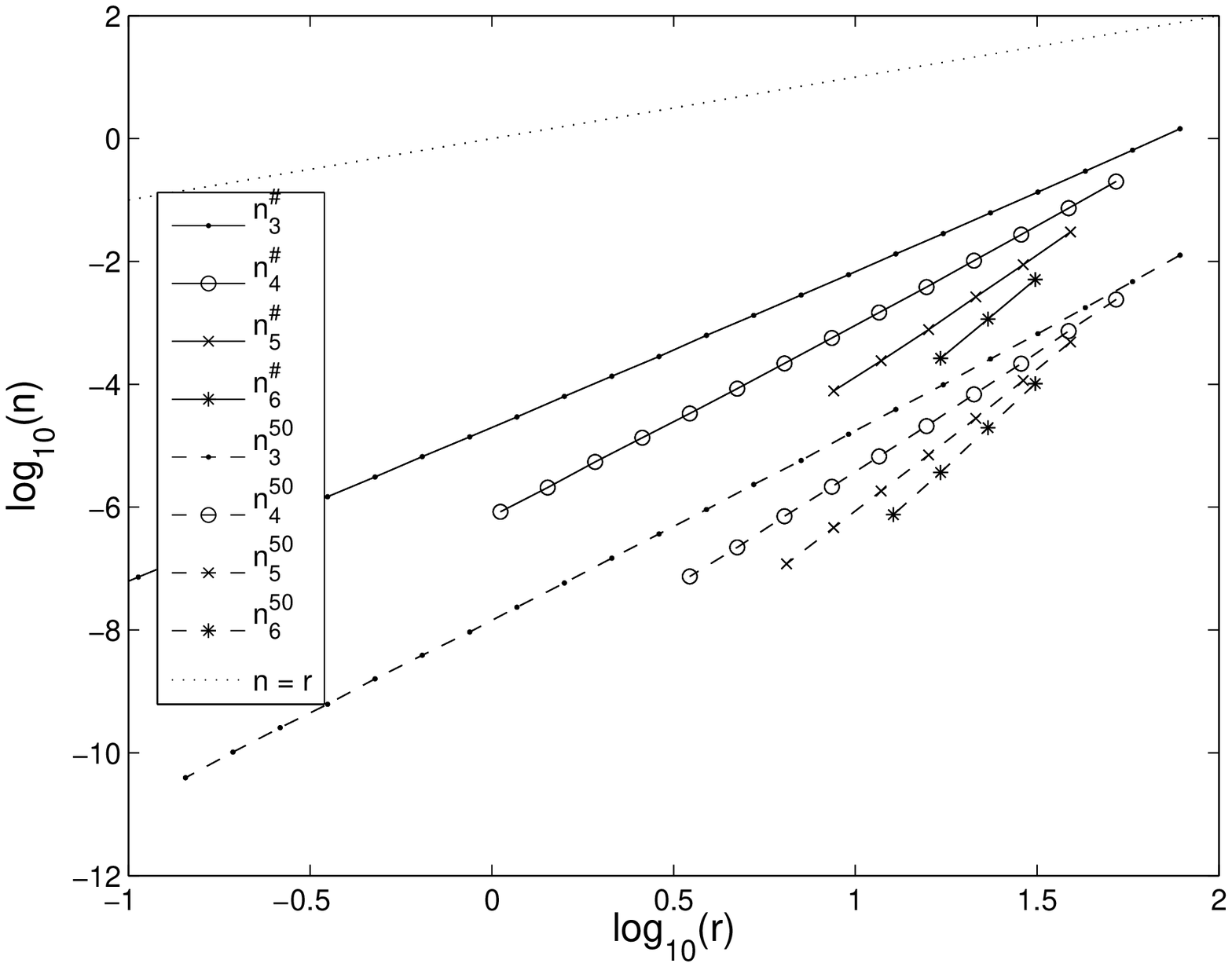,width=7.5cm}
  \caption{{\bf Estimates of nucleation rates from stochastic
      simulations.}    $n^\#_k$ and $n^{50}_k$ vs. $r_k$ for
    $k=3,\;4,\;5$ and $6$. $\epsilon = 0.1$.
The line $n=r$ is
      also plotted to illustrate values of $n$ if there were no
      improvement in nucleation rates with assembly slowdown.}
\label{instantaneous_nucleation}
\end{figure}

Having proven in the previous section that zig-zag tile sets can be
designed to achieve arbitrarily low spurious nucleation rates relative
to the growth rates (using a loose upper bound), we now ask
whether the nucleation barrier provided by zig-zag tile sets is
sufficient for practical implementation in the laboratory (which
requires more accurate quantitative assessments).  There are two main
concerns: first, as each tile must be synthesized, $k$ must be small
(6 is currently practical, while 50 is currently too large); second,
assembly time must not be too long. (Growing 1000 layers of seeded
crystals with less than 1\% spurious nucleation---which we refer to as
the ``typical reaction''---seems like a reasonable goal to accomplish
within one week.)  Because the analytic bounds of
Section~\ref{asymptotic_analysis} are too loose to allow us to obtain
a realistic evaluation of nucleation rates, we now develop more
accurate numerical calculations and stochastic simulations for
estimating spurious nucleation rates.

  The analysis in Section~\ref{asymptotic_analysis} overestimates the
  spurious nucleation rate in three ways.  First, it overestimates the
  concentration of almost all kinds of assemblies by assuming they
  have the same concentration as a rectangular assembly of the same
  length and width, and it overcounts the number of different types of
  assemblies.  Second,
  Lemma~\ref{powered_accretion_steady_state_bound_lemma} shows that
  the spurious nucleation rate at steady state is the maximal spurious
  nucleation rate.  However, it may take longer to approach steady
  state than the time needed to run a ``typical reaction,'' and far
  from steady state, the spurious nucleation rate may be much smaller
  than the spurious nucleation rate at steady state.  Lastly, this
  analysis defines a spurious nucleation event for a zig-zag tile set
  of width $k$ as a reaction that produces an assembly of width $k$,
  and neglects the backward reaction.  In practice, many reactions
  that form an assembly of width $k$ are unfavorable, so that the
  product assembly frequently shrinks back to a sub-critical size
  instead of growing larger.  When conditions only slightly favor
  growth, even assemblies containing several layers have a reasonable
  chance of shrinking to nothing before they grow substantially.
  We expect $n_k << n^+_k$ in this case.

  While it is not possible to compute the nucleation rate exactly, in
  this section we describe three numerical techniques that correct
  each inaccuracy described above for zig-zag tile sets of widths
  $k=3,\;4,\;5 $ and $6$.  In Section~\ref{steady_state_section}, we
  compute the rate at
  which ribbons of width $k$ are formed at steady state using a
    much more accurate count of the number and steady state
    concentration of assemblies..  These computations show that
  the analytic bound of Theorem~\ref{powered_accretion_theorem} is too
  high by at least 4 orders of magnitude for the range of parameters
  studied.
  In Section~\ref{instantaneous_nucleation_section} we use a
  stochastic simulation of tile assembly to estimate the rate of
  spurious nucleation reactions $R^{in}_k$.  Our results indicate that
  spurious nucleation reactions occur during a ``typical reaction'' at
  a rate that is no more than an
    order of magnitude lower than the rate at steady state
computed in Section~\ref{steady_state_section}.  In
  Section~\ref{large_assemblies_section}, we use the stochastic
  simulation to investigate whether the rate of spurious nucleation
  reactions ($n^+_k = F(R^{in}_k,s)$) in a typical reaction accurately
  predicts the rate at which large assemblies appear (which at steady
  state is equivalent to $n_k(s) = F(R^{in}_k,s) - F(R^{out},s)$).  We
  find that for the range of parameters studied, at least 99\% of
  assemblies that reach full width will melt before growing into large
  crystals, and thus our other estimates of spurious nucleation rates
  may be overestimates of $n_k$ by at least two orders of magnitude.
  In Section~\ref{other_considerations}, we show that these results
  together indicate that a zig-zag tile set of width 5 or 6 should be
  large enough to prevent almost all spurious nucleation in a
  ``typical reaction'', while maintaining reasonable assembly speeds.
  We conclude with an important caveat to these results.  Our results
  are derived under a powered accretion model of kTAM, while in
  experiments, small assemblies may aggregate rather than growing
  exclusively by single tile additions, thus potentially producing
  nuclei that reach a critical size more quickly than our simulations
  indicate.

\subsection{Spurious Nucleation Rates at Steady State}
\label{steady_state_section}
Recall that for a zig-zag tile set of width $k > 2$, the steady state rate of spurious
nucleation reactions is given by the sum
\begin{equation*}
n^+_k = \lim_{s \rightarrow \infty} F(R^{in}_k,s) = \sum_{l=1}^{\infty} \ \ \ 
 \sum_{\substack{A + t\rightarrow B + \mathit{t}  \in R^{in}_k \\ \text{s.t. } length(A)=l}} k_f[A]_{ss} e^{-G_{mc}}, 
\label{sn_recap}
\end{equation*}
which ignores the rate at which
spuriously nucleated assemblies dissolve back into pre-nucleated
assemblies.  While $[A]_{ss}$ is known (at
steady state, for an assembly $A$ with $n$ tiles and $b$ bonds,
$[A]_{ss} = e^{bG_{se} - nG_{mc}}$), 
it is not practical to compute the sum exactly
because there are an infinite number of spurious nucleation reactions.
Additionally, it can be impractical to evaluate the inner sum even for
a single value of $l$: no efficient algorithm is known  for exactly enumerating the reactions in $R^{in}_k$
(see e.g. \cite{Golomb1994} for the related problem of counting
  polyominos).  The number of distinct reactions increases
exponentially with the length of $A$, so it is prohibitive
to calculate all but the first terms of the sum.

Despite these difficulties, the expression can be calculated
precisely, with known error bounds, for many $k$.  The following lemma
shows that under many reaction conditions of interest, the sum
converges quickly, and its value can be
approximated by summing only the first few terms:
\\
\begin{lemma}
  When $G_{se} > (\ln 10)(k-2) + \ln{4}$, \( G_{mc} = \: 2G_{se} -
  \epsilon\), \( 0 \le \epsilon < \frac{1}{2k-3} \) $, k > 2$
  and $l$ is even,
\begin{equation*}
\sum_{p=l+1}^{\infty} \left( \sum_{\substack{A + t\rightarrow B + \mathit{t}  \in R^{in}_k\\ \text{s.t. } length(A)=p}} k_f[A]_{ss}e^{-G_{mc}} \right) < 
2 \left( \sum_{\substack{A + t\rightarrow B + \mathit{t}  \in R^{in}_k\\ \text{s.t. } length(A)=l}}
k_f[A]_{ss}e^{-G_{mc}} \right)
\end{equation*}
\label{fast_convergence_lemma}
\end{lemma}
\\
\begin{proof}
See Appendix \ref{fast_convergence_lemma_proof}. 
\end{proof}\\

Thus, to calculate the spurious nucleation rate up an accuracy of
$\frac{1}{\varepsilon}$, it is only
necessary to compute the inner sums of the series until the sum the
current value of $l$ is (even and) less than
$\frac{1}{2\varepsilon}$.  (Note that
this approach does not directly yield a proof of an
analytic bound for arbitrary $k$, because the
formula for the nucleation rate is not a closed form expression.)

We have used this series truncation method to calculate the rate of
spurious nucleation to 1 part in $10^4$
for $k = 3,\;4,\;5$ and 6 and for a range of $G_{se}$, $G_{mc}$ for
which $\epsilon = 0.1$.  The values of $G_{se}$, $G_{mc}$ and $k$ used
were in a regime in which Lemma~\ref{fast_convergence_lemma} applies.
The results are shown in Figure~\ref{steady_state_nucleation}.

\begin{figure}
\centering
\epsfig{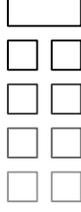}
\caption{{\bf Hypothesized principal critical nucleus for most
    spurious nucleation reactions.} The rate of spurious nucleation
  reactions by this assembly (shown in
  successively lighter shades of gray for tile
  sets of widths $3,\;4,\;5$ and 6) accounts for a large portion of spurious
  nucleation at slow speeds, and also accounts for the rate of
  increase in spurious nucleation rates as assembly gets faster.}
\label{critical_nucleus}
\end{figure}

In addition to the numerical calculations providing lower estimates,
the slopes of $\log n^+_k$ vs $\log r_k$ in
Figure~\ref{steady_state_nucleation} are larger than
those of $\log n^B_k$ vs $\log r_k$.
Specifically the numerical calculations give slopes $\frac{k+2}{2}$,
compared to the analytic bounds that give slopes
$\frac{k}{2}$.  Is this reasonable?  In the limit as $G_{mc}
\rightarrow \infty$, all spurious nucleation should be dominated by
the single species with the highest steady state concentration (adding
tiles become so unfavorable that other species can be neglected.)  The
analysis in Section~\ref{zig_zag_growth_and_nucleation} suggests that
this assembly is the one shown in Figure~\ref{critical_nucleus}.  The
steady state concentration of this assembly $A$ for a tile set of
width $k$ is $[A]_{ss} = e^{-(2k-3)G_{mc} + (3k-6)G_{se}} =
e^{-kG_{se}+(2k-3)\epsilon}$, where $\epsilon = 2G_{se} -
  G_{mc}$.  If all forward nucleation reactions involve $A$, then
\begin{equation}
 n^+_k \approx 2 k_f [A]_{ss} e^{-G_{mc}} = 2 k_f e^{-(k+2)G_{se}+(2k-4)\epsilon}
\label{n_k_plus_estimate}
\end{equation}  
while the speed of growth is
\begin{equation*}
 r_k =  \frac{k_f}{k-1} \left(e^{-G_{mc}} - e^{-2G_{se}}\right) = k_f  e^{-2G_{se}} \frac{ e^\epsilon - 1} {k-1},
\end{equation*}
and thus the slope would be $\frac{k+2}{2}$, as observed.  The rough estimate of $n^+_k$ given
  in Equation~\ref{n_k_plus_estimate} is within a factor of three of
the value calculated to an
  accuracy of 1 part in $10^4$, as guaranteed by
  Lemma~\ref{fast_convergence_lemma}.  

\subsection{Stochastic simulations for estimating forward nucleation rates before steady state is achieved}
\label{instantaneous_nucleation_section}

\begin{table}[t]
\begin{center}
\begin{tabular}{|p{8ex}|p{13ex}|p{13ex}|p{13ex}|}
  \hline 
  Tile set width & $n^+_k$ & $n^\#_k$ & $n^{50}_k$ \\
  \hline
  3 & $3 \times 10^{24}$ years & $1 \times 10^{23}$ years& $9 \times 10^{5}$ years\\
  \hline 
  4 & $2 \times 10^8$ years & $3 \times 10^6$ years&  $7$ years\\
  \hline
  5 & 900 years & $10$ years&  $20$ days\\
  \hline  
  6 & 2 years &  $20$ days&  $20$ hours\\
  \hline 
\end{tabular}
\caption{{\bf Estimated time needed to grow 1000 layers in 1 pmol of free tiles without
    any spurious nucleation, based on approximations of the spurious nucleation rate.} While the total number
    of tiles in each reaction is constant, the volume differs
    depending on the speed, which determines the
    concentration and the spurious nucleation rate.}
\label{time_needed_table}
\end{center}
\end{table}

\cut{ Rebecca's numbers:
\begin{table}[t] \begin{center}
    \begin{tabular}{|p{8ex}|p{13ex}|p{13ex}|p{13ex}|} 
\hline 
  Tile set width & $n^+_k$ & $n^\#_k$ & $n^{50}_k$ \\
 \hline 
3 &$90,000$ years & $40,000$ years & 9 hours\\ 
\hline 
4 &10 days & 5 days & 20 minutes \\
\hline 
5 &30 hours & 1 hour & 4 minutes\\ 
\hline 
6 &20 minutes & 10 minutes & 2 minutes \\ 
\hline 
\end{tabular}
    \caption{{\bf Estimated time needed to grow 1000 layers such that less than
        1 percent of assemblies are spuriously nucleated based on approximations of the spurious nucleation rate.} The seed concentration is $\frac{1}{1000}$ the concentration of free tiles $e^{-G_{mc}}$.}
\label{one_percent_time_needed_table}
\end{center}
\end{table}
}

\begin{table}[t] \begin{center}
    \begin{tabular}{|p{8ex}|p{13ex}|p{13ex}|p{13ex}|} 
\hline 
  Tile set width & $n^+_k$ & $n^\#_k$ & $n^{50}_k$ \\
 \hline 
3 &$1 \times 10^{11}$ years & $1 \times 10^{10}$ years & 200 days\\ 
\hline 
4 &40 years & 5 years & 12 hours \\
\hline 
5 &10 days & 2 days & 1 hour\\ 
\hline 
6 &7 hours & 2 hours & 15 minutes \\ 
\hline 
\end{tabular}
    \caption{{\bf Estimated time needed to grow 1000 layers such that less than
        1 percent of assemblies are spuriously nucleated based on approximations of the spurious nucleation rate.} The seed concentration is $\frac{1}{1000}$ the concentration of free tiles $e^{-G_{mc}}$.}
\label{one_percent_time_needed_table}
\end{center}
\end{table}

In order to determine whether the steady state approximation is accurate
  over a typical spurious nucleation reaction, we simulated zig-zag
tile assembly for tile sets of widths
$k=3,\;4,\;5$ and $6$ and measured the rates of
spurious nucleation events during the time it should take to grow 1000
layers from seeds.  Since there are an infinite number of powered
accretion reactions, exact simulation of growth under the kTAM using
mass action dynamics is not possible.  Instead, we simulated assembly
growth using stochastic chemical reaction dynamics for discrete
  numbers of molecular assemblies.  Here, simulation is
  possible because even though the probability of any of the infinite
  number of species arising is larger than 0, the total number species
  tracked at a given time is finite.  To approximate the nucleation
rate, we simulated a tiny reaction volume, and used these
results to predict the nucleation rate in a much larger volume.

We used the Gillespie algorithm~\cite{Gillespie76} to sample the
trajectories of stochastic dynamics of the zig-zag tiles in a small
volume $V$, whose value is chosen to ensure the accuracy of our
nucleation rate estimate as described below.  Following the powered
model, our simulation assumes the concentration of each tile type to
be constant and explicitly tracks each assembly in the volume
containing more than one tile.  Initially, no multi-tile assemblies
are present.  Single tiles are present at a concentration of
$e^{-G_{mc}}$, so the rate of two tiles colliding (and thus
producing a new assembly to be explicitly tracked) is $\mathbf{A}k_f V
e^{-2G_{mc}}$ molecules / second where $\mathbf{A}$ is Avogadro's
number.
For each assembly containing two or more tiles, the rate of tile
addition at each available site is $k_f e^{-G_{mc}}$ and the
rate at which a tile with $b$ bonds falls off an
assembly is $k_f e^{-bG_{se}}$.

For $k=3,\;4,\;5$ and $6$ and a range of $G_{se}$ and $G_{mc}$ where
$\epsilon = 0.1$, we counted the number of spurious nucleation events,
$m$, that took place over the time course of a ``typical reaction'',
$s = 1000/r_k$, in a volume $V$ that was chosen large enough to ensure
that statistical error in $m$ is less than 10 percent of its value ($P
> 0.95$)\footnote{That is, twice the standard deviation of the number
  of nucleation events per simulations is less than ten percent of the
  average number of nucleation events per simulation.}.  If our
simulations yield a nucleation rate of $m$ events per second, the
molar rate of nucleation events for a bulk volume is given by
$n^\#_k \approx
  \frac{m}{V \mathbf{A}}$.  The results of the simulation---which
were possible only for small enough $G_{se}$ such that nucleation
events were frequent enough to be counted---are shown in
Figure~\ref{instantaneous_nucleation}.  For $k=3$ and $k=4$, these
rates are within a factor of 2 of the linear extrapolation of the
curves from Figure~\ref{steady_state_nucleation}, and for $k=5$ and
$k=6$ these rates are within a factor of 10, indicating that the
choice in Section~\ref{asymptotic_analysis} to bound nucleation rates
based on steady state concentrations did not affect our estimate of
nucleation rates too greatly.  This should be expected, given that
under the conditions we studied most steady state nucleation appears
to involve assemblies like the one shown in
Figure~\ref{critical_nucleus}.

\subsection{Nucleation of Long Ribbons}
\label{large_assemblies_section}
In this paper, we have defined a spurious nucleation reaction for a
zig-zag tile set of width $k$ as a reaction in which an assembly of
width $k-1$ grows to width $k$.  The goal was that this definition
would be inclusive, such that all long ribbons would undergo at
least one spurious nucleation reaction, but not too loose, such that
most spurious nucleation reactions lead to a long ribbon.  However,
many of these spurious nucleation reactions are not energetically
favorable---an assembly may briefly reach width $k$ before a tile
falls off. The assembly then either melts or undergoes another
spurious nucleation reaction.

At what rate do long ribbons appear?  Using the stochastic simulation
described in the last section and the same range of physical reaction
parameters, we measured $m'$, the number of ribbons containing 50
tiles or more that were present at the end of a ``typical reaction'',
for the widths 3, 4, 5, or 6.  $m'$ is an
  estimate of the number of spurious nucleation events that did {\it
  not} subsequently melt, and thus it provides the basis for an
estimate for $n_k$.  As only those crystals that nucleated
sufficiently far before the end of the simulation will have grown to a
large enough size to have been counted,  we use the formula
$n^{50}_k \stackrel{def}{=} \frac{m'} {(s -
    \left(50r_k / (k-1)\right)) V \mathbf{A}}$, where $s$ is the time
  of the simulation and $50r_k / (k-1)$ is the approximate time to
  grow to 50 tiles via zig-zag growth.  The results are shown in
Figure~\ref{instantaneous_nucleation}.

These simulations suggest that much of the looseness of the
analytical bound on nucleation, $n^B_k$, is caused by our neglecting crystals which under go a
nucleation reaction and subsequently melt.  These simulations suggest
that at least 99\% of crystals that undergo a spurious nucleation do
not grow into crystals consisting of 50 or more tiles.  That is, $n_k
\ll n^+_k$.

\subsection{Expected Effectiveness in Practice}
\label{other_considerations}

Do these results indicate that nucleation control with tile sets of
width 6 or less are good enough?  Recall that our ``reasonable goal
for a typical reaction'' addresses how much time is needed to grow
seeded ribbons of 1000 layers with less than 1\% of the crystals being
spuriously nucleated.  The fraction of crystals that are spuriously
nucleated is given by $f = \frac{L}{c} \frac{n_k}{r_k}$, where $L$ is
the number of layers to be grown on seeds, and $c$ is the
concentration of seeds.  While the simulations only measured $n_k$ for
large values of $r_k$, it is possible to approximate $n_k$ for smaller
values of $r_k$ by assuming that the graph of $\log(n_k)$
vs. $\log(r_k)$ continues as a line with constant slope as $r_k$ and
$n_k$ decrease.  
We consider two cases.  The first situation (more stringent than ``reasonable'') is to grow many ribbons --- say, from 1~picomole of each tile type, making $6\times10^{8}$ ribbons of length 1000 --- and not have more than a single spuriously nucleated crystal, i.e., $f  < 1.67\times10^{-9}$ and $c = e^{-G_{mc}}/1000$.  
To satisfy this constraint, we express $f$ in terms of $c$  using our estimates for $n_k$, and solve for $c$ (and hence the concentration of free tiles and the volume of the reaction by $\mathbf{A} e^{-G_{mc}} V = 10^{-12}$ moles).  
The time needed to grow the crystals is therefore given as $L/r_k$ for these conditions.   
The results, shown in Table~\ref{time_needed_table}, suggest that if $n^{50}_k$ is accurate, then this stringent goal could be met using a width 6 zig-zag tile set and a day of growth.  The looser estimates and smaller tile set widths are less encouraging.
The second situation we consider is the ``reasonable'' one; again, $c = e^{-G_{mc}}/1000$, but now we only require $f < 0.01$.  The results are shown in Table~\ref{one_percent_time_needed_table}.   In this case, acceptable growth fidelity is predicted to be achieved in less than an hour for width 6, and for only slightly longer times for widths 5 and 4.
However, all these estimates are very sensitive to the coefficients of the linear fits to $\log{n_k}$ vs $\log{r_k}$, which are imperfect because the relationship is not perfectly linear.
\cut{
The results, shown in
Tables~\ref{time_needed_table}~and~\ref{one_percent_time_needed_table},
suggest that by using a zig-zag tile set of width 6, just a few hours
would be enough to avoid {\it all} spurious nucleation in a mixture
containing 1 pmol of each tile type.  If a small amount of spurious
nucleation is acceptable, such that the concentration of spuriously
nucleated assemblies comprised less than 1\% of the total, assembly of
1000 layer crystals can be complete in just a few minutes.  However,
in both these cases it is important to note that predictions for the
times required are made based on the assumption that the slope
$\log{n_k}$ vs $\log{r_k}$ is constant, which is not strictly true.
}

The analysis and simulations in this section support the idea that
nucleation control using the zig-zag tile set not only works in
  theory, but should be practical.  While in most
respects our models appear complete, two effects which may be
important in the actual process of assembly are not included.  One
such effect is tile depletion: while our model
considers the concentration of free tiles to be constant, in a typical
experiment tiles are used up because they join assemblies.  Since the
rate of spurious nucleation is concentration dependent, we would
expect the rate of spurious nucleation to be larger at the beginning
of a reaction, when almost all free tiles remain, than at the end,
when many tiles are used up.  Because of this effect, our simulations
may actually {\it overestimate} the spurious nucleation rate in
  experimental systems.

However, our simulations also neglect an important possible reaction
pathway that may greatly increase the rate of spurious nucleation.
While our model assumes tiles must be added to assemblies one at a
time, in an experiment, small assemblies can also attach to each
other.  The formation and joining of several small assemblies may be
faster than the spurious nucleation pathways described in this paper.
A complete understanding of spurious nucleation of zig-zag tiles
requires an understanding of the speed of spurious nucleation
reactions caused by the joining of small
  assemblies.

\section{Conclusions}

\subsection{Nucleation of Algorithmic Self-Assembly} 

Our original motivation  was to show that
self-assembly programs that work in the aTAM, in which it is
straightforward to design tile sets that algorithmically assemble any
computationally defined structure, can also be made to work in the
more realistic kTAM.

While tile sets that assemble correctly via unseeded growth in the
aTAM with a threshold of $\tau=1$ will assemble correctly in the kTAM
under the right conditions, programs to assemble structures can be
exponentially larger (in terms of number of tile types) than those
with a threshold of $\tau=2$~\cite{RothemundWinfree2000}. However,
tile sets that are designed to assemble via seeded growth in the aTAM
with a threshold $\tau=2$ may fail in the kTAM because mismatch, facet
and spurious nucleation errors occur.  These problems are ameliorated
in the limit of slow assembly speed~\cite{Winfree1998b}.  Other work
has described methods to control mismatch errors and facet errors
without significant
slowdown~\cite{WinfreeBekbolatov2003,Chen_Goel2004,Reif_Sahu_Yin2004}.
Here, we have developed a construction that may be
  used to correct the last discrepancy, spurious nucleation errors,
again without significant slowdown.

It remains to be formally proven that these constructions can be
combined to control all types of errors simultaneously for any tile
set of interest.  No major difficulties are expected, however, in
large part because mismatch and facet errors can both be controlled by
a single mechanism~\cite{Chen_Goel2004} and the control of spurious
nucleation errors works independently of this mechanism. Both methods
work by transforming an original tile set which works in the aTAM at
$\tau=2$ into a new (typically larger) tile set that is more robust to
particular kinds of errors in the kTAM. 

After this paper was submitted, experimental demonstration of a
decrease in nucleation rates with ribbon width have supported the
predictions made here~\cite{SchulmanWinfree2007}.  Further experiments
combining the techniques described here with proofreading techniques,
as predicted, resulted in algorithmic assembly where both mismatch
error rates and spurious nucleation error rates are
low~\cite{Barishetal2009}, and has enabled other algoithmic
self-assembly experiments~\cite{Fujibayashietal2008}.  It remains to
be seen whether facet nucleation rates can be lowered in experimental
demonstrations of algorithmic self-assembly, but the principal
mechanism in the theoretical proposal for lowering facet nucleation
error rates~\cite{Chen_Goel2004} has been has been experimentally
confirmed~\cite{Chenetal2007}.

\subsection{Detection of a Single DNA Molecule}

\begin{figure}[t]
\centering
\epsfig{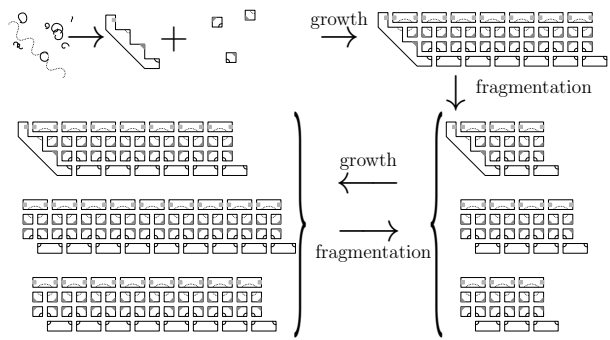}     
\caption{ {\bf Exponential amplification of assemblies.}  Probe
  strands assemble onto a target sequence to create a seed assembly,
  which nucleates zig-zag growth.  Periodic fluid shear causes
  fragmentation of zig-zag assemblies, leading to exponential
  amplification.  The diagonal structure of the seed assembly shown
  here is a natural shape for assembling tiles on a scaffold
  strand~\cite{Rothemund2004sier}. }
\label{amplified_nucleation}   
\end{figure}

Control over nucleation in algorithmic self-assembly can be seen as a
special case of the detection of a single molecule.  For a tile set of
sufficiently large width, essentially nothing happens when no seed
tiles are present, whereas if even a {\em single} seed tile is added,
growth by self-assembly will result in a macroscopic assembly.
Theorem~1 shows that the {\em false-positive} rate for detection can
be made arbitrarily small by design; the {\em false-negative} rate in
the kTAM is approximately 0.  Although this idealized model does not
consider many factors that could lead to poorer detection in a real
system, we don't know of any insurmountable problems with implementing
single-molecule detection this way.  So far, experiments have shown
that seeded growth can be much faster than unseeded
growth, even when seeds are present at much lower concentrations than
the elements of the zig-zag tile
set~\cite{SchulmanWinfree2007,Barishetal2009}, and no lower limit for
detection with current technology has been established.

There are, however, two immediate drawbacks.  First, detecting
seed-tile assemblies is not as useful as detecting arbitrary DNA
sequences. Second, the linear growth of a single zig-zag assembly
would require a long time lapse before a macroscopic change is
perceptible. As sketched in Figure~\ref{amplified_nucleation},  both obstacles appear surmountable.  First, as
in~\cite{Mao2000,Yan2003a}, a set of strands can be designed to
assemble double-crossover molecules on a (sufficiently long) target
strand with nearly arbitrary sequence, thus creating the seed assembly
if and only if the target strand exists.  Second, since fluid shear
forces can fragment large DNA
assemblies~\cite{HariadiYurke2009}, intermittent application of these forces could break
large zig-zag assemblies, increasing the number of growing ends with
each fragmentation episode.  This fragmentation process can be
expected to lead to exponential growth in the number of zig-zag
assemblies without increasing the false-positive rate.  (When a
spuriously nucleated assembly does eventually form, of course, it will
also be exponentially amplified.)

Based on the analyses of the previous section, we can estimate the
effectiveness of this procedure.  Is there a reasonable tile set width
for which a single seed could amplify to a level of detectability in a
reasonably short time without any spurious nucleation occurring within
the given volume?  Specifically, given a 10 $\mu$L reaction volume, a
minimum detection level of $10^5$ crystals and a protocol in which
assemblies split after growing on average to size 200 layers, we would
like to determine the minimum time and tile set width that meet
these requirements.  Creating $10^5$ crystals requires first growing
from the seed to size 200, then 17 cycles of fragmentation followed by
growing 100 additional layers (50 on each side), so amplification
requires $t_a = 1050/r_k$ seconds.  The expected time
for the first nucleation event is $t_f = \frac{1}{n_kV\mathbf{A}}$,
and our criteria for reliable detection is $t_f > 100t_a$, i.e.,
$\frac{n_k}{r_k} <
  \frac{1}{105000V\mathbf{A}}$.  Based on the $n_k^{50}$
  estimates described in
  Section~\ref{instantaneous_nucleation_section},
we use the approximation $\log(n_k) + 2 =
\frac{k+2}{2}(\log(r_k)-1.7)$.  Solving for $t_a$ as a function of
$k$, we find that good results are obtained for experimentally
feasible widths.  For example, with $k=12$, reliable detection of a
single seed in $V=10\;\mu$L (i.e. $0.13$ attomolar
  concentration) is $t_a \approx$ 26 hours.

\subsection{Exponential Replication of Inheritable Information} 

The zig-zag constructions detailed in this paper propagate a single
bit of information: the presence or absence of the seed tile. Using a
tile set that simply copies information, we could use the exponential
amplification reaction to detect and identify one of several different
target strands, by creating a tile set where the seed assemblies for
each target strand contain a different pattern of 1s and 0s. 

Furthermore, considering the amplification process as replication, the
information encoded in the strip's width can be seen as a form of
inheritable information~\cite{SchulmanWinfreeECAL2005}, related to
Graham Cairns-Smith's proposal for information replication within
clays~\cite{CairnsSmith71,CairnsSmith1988}.  A zig-zag assembly
replicates (in the appropriate culture medium consisting of tiles) by
growth of new layers followed by random
fission~\cite{GregEganWangsCarpets}.  Errors during growth, bit flips,
as well as errors that increase or decrease the width of the assembly,
are inherited.  If one sequence of tiles has a greater reproductive
fitness than other sequences---for example, by having a different
growth or fission rate---then natural evolution can be expected to
occur.  In principle, the right selective pressures on such a process
could induce the formation of arbitrarily complex crystal
genotypes~\cite{SchulmanWinfree2008}.

\section{Acknowledgements}

The authors are grateful to Ho-Lin Chen, Ashish Goel, Zhen-Gang Wang
and Deborah Fygenson for helpful advice and discussions and to Donald
Cohen for advice on extending the results included here. 
\appendix

\section{Mass-action kTAM satisfies Detailed Balance}
\label{detailed_balance_proof}

This section contains proof that the mass-action kTAM used in this
paper satisfies detailed balance within $\mathcal{A}_{2+}$.  We prove two facts
necessary to show this.

The proof also applies to the case, not considered in this paper,
where different tile types have different (but constant)
concentrations.  For a tile \(\mathit{t}\), \( S(\mathit{t})\) is
defined as the relative concentration of its corresponding tile type.
Unit concentration is $e^{-G_{mc}}$, such that the concentration of
tile type $\mathbf{t}$, \([\mathbf{t}] = S(\mathbf{t}) e^{-G_{mc}}\).
Additionally, while only equal strength sticky ends are considered in
this work, this proof shows that detailed balance applies to a model
of self-assembly with arbitrary sticky end strengths.

\begin{lemma}
  For all reaction pairs $A + t \rightarrow B + t$ and $B \rightarrow
  A$, \( k_f[t] [A]_{ss} = k_r[B]_{ss} \), where $k_f$ and $k_r$ are
  the rates of the respective reactions. 
\end{lemma}

\begin{proof}
\begin{align*}
k_r[B]_{ss} &= k_f e^{G^{\circ}(B) - G^{\circ}(A)}[B]_{ss} \\
           &= k_f e^{G^{\circ}(B) - G^{\circ}(A)}e^{-G(B)} \\
           &= k_f e^{G^{\circ}(B) - G^{\circ}(A)}e^{-\left(G^{\circ}(B) + \left(\sum_{t' \in B} G_{mc} - \ln(S(t'))\right)\right)} \\
           &= k_f e^{- G^{\circ}(A)}e^{-\sum_{t' \in B} \left(G_{mc} - \ln(S(t'))\right)}.
\end{align*}
Because $A + t = B$, 
\begin{align*}
           &= k_f e^{- G^{\circ}(A)}e^{-\sum_{t' \in A} \left(G_{mc} - \ln(S(t'))\right)}e^{-G_{mc} + \ln(S(t))} \\ 
           &= k_f e^{- G(A)}e^{-G_{mc} + \ln(S(t))} \\ 
           &= k_f [A]_{ss} [t].
\end{align*}

\end{proof}

\begin{lemma}
For reaction pairs $t_1 + t_2 \rightarrow A$ and $A \rightarrow
\emptyset$, $k_f[t_1][t_2] = k_r[A]_{ss}$. 
\end{lemma}

\begin{proof}
\begin{align*}
k_r[A]_{ss} &= k_f e^{G^{\circ}(A)} [A]_{ss} \\
            &= k_f e^{G^{\circ}(A)} e^{-G(A)} \\
            &= k_f e^{G^{\circ}(A)} e^{-\left(G^{\circ}(A) + 2G_{mc} - \ln(S(t_1)) - \ln(S(t_2))\right)} \\
            &= k_f e^{-G_{mc} + \ln(S(t_1))} e^{-G_{mc} + \ln(S(t_2))} \\
            &= k_f [t_1][t_2].
\end{align*}
\end{proof}

\section{Steady State Concentration as a Bound on Assembly 
Concentration in a Powered Accretion Self-Assembly Model}
\label{steady_state_bound_proof}
This section contains the proof of Lemma
\ref{powered_accretion_steady_state_bound_lemma}: In a mass-action powered accretion kTAM, if in
  the initial state only single tiles have a positive concentration,
  then every assembly has a concentration less than or equal to its
  steady state concentration at all time points.

Suppose that this lemma is not true.  Then there is a time at which
the concentrations of one or more assemblies exceed their values at
steady state.  Since the concentrations of all assemblies are zero
initially, there must be a first time point $s$ at which for
at least one assembly $A$, $[A]=[A]_{ss}$.  At this
time point, the concentrations of all other assemblies are
either at or below their respective steady state concentrations.
The rate of
change of $[A]$ is given by Equation~\ref{dA_ds_equation}:

\begin{multline*}  
\frac{d[A]}{ds} =k_f \biggl(\sum_{\substack{A + \mathit{t} \rightarrow
B + \mathit{t},\\
B \rightarrow A \; \in R}} e^{G^{\circ}(B) - G^{\circ}(A)}[B] -
[A]e^{-G_{mc}} \;+ \\
 \sum_{\substack{B + \mathit{t} \rightarrow A + \mathit{t},\\ 
A \rightarrow B \; \in R}} [B]e^{-G_{mc}} -
e^{G^{\circ}(A) - G^{\circ}(B)}[A] \;+  \sum_{\substack{\mathit{t_1} +
\mathit{t_2} \rightarrow A+\mathit{t_1}+\mathit{t_2},\\
A \rightarrow \emptyset \in
R}}e^{-2G_{mc}} -
e^{G^{\circ}(A)}[A] \biggr). 
\end{multline*}

Consider a single term in the second summation, \(
[B]e^{-G_{mc}} - e^{G^{\circ}(A) - G^{\circ}(B)}[A] \),
involving some assembly $B$.  We know that $[A]$ has reached its
steady state concentration, so \( [A] = e^{-G(A)} \). By assumption, \(
[B] \le [B]_{ss} = e^{-G(B)} \).  Assembly $A$ includes one more tile,
$t$, than does assembly $B$, so \( G^{\circ}(A) - G^{\circ}(B) = G(A)
- G(B) - G_{mc}\).  Therefore,

\begin{align*}
[B]e^{-G_{mc}} - e^{G^{\circ}(A) - G^{\circ}(B)}[A]  &=
[B]e^{-G_{mc}} - e^{G(A) - G(B) - G_{mc}} [A] \\ 
&= [B]e^{-G_{mc}} - e^{G(A) - G(B) - G_{mc}} e^{-G(A)} \\ 
&= [B]e^{-G_{mc}} - e^{-G(B)}e^{-G_{mc}} \\
&= e^{-G_{mc}} \left([B] - e^{-G(B)}\right)\\ &\le 0. 
\end{align*}

Similarly, for an assembly $B$ that is a term in the first summation,
$B$ has the extra tile $t$ so that $G^{\circ}(B) - G^{\circ}(A) = G(B)
- G(A) - G_{mc}$.  The term can be simplified to

\begin{align*}
e^{G^{\circ}(B) - G^{\circ}(A)} [B] - [A]e^{-G_{mc}} 
&= e^{G(B) - G(A) - G_{mc}} [B] - e^{-G(A)}
e^{-G_{mc}}\\ 
&\le e^{G(B) - G(A) - G_{mc}} e^{-G(B)} -
e^{-G(A)} e^{-G_{mc}}\\
&= 0. 
\end{align*}

The terms in the third summation are also non-positive, since
\begin{align*}
e^{-2G_{mc}} - e^{G^{\circ}(A)}[A] 
&= e^{-2G_{mc}} - e^{G^{\circ}(A)}e^{-G(A)}\\
&= e^{-2G_{mc}} - e^{G(A) -2G_{mc}}e^{-G(A)}\\ 
&= 0. 
\end{align*}
The change in concentration \(\frac{d[A]}{ds}(s)\) is composed
entirely of terms of this form. Since each of these terms is
non-positive, \(\frac{d[A]}{ds}(s)\) is non-positive when
$[A]=[A]_{ss}$.  Thus, $[A]$ can never rise above its
steady state value.

As in Appendix~\ref{detailed_balance_proof}, this proof also applies
to a model of self-assembly with arbitrary stoichiometry and sticky
end strengths.


\section{Fast Convergence of Nucleation Rates at Steady State}
\label{fast_convergence_lemma_proof}

This section contains a proof of Lemma~\ref{fast_convergence_lemma}
for zig-zag tile sets of width $k$.  

We start by re-writing the
lemma to use convenient notation to refer to the inner sums within the
series for $n_k^+$, which refer to the rate of spurious nucleation
events involving assemblies $A$ of width $k-1$ and length $l$:
$$N_p = \sum_{\substack{A + t\rightarrow B + \mathit{t}  \in R^{in}_k\\ \text{s.t. } length(A)=p}} k_f[A]_{ss}e^{-G_{mc}}, $$
such that $n_k^+ = \sum_{l=1}^\infty N_l$.  Now, Lemma~\ref{fast_convergence_lemma} may be stated as:
\vspace{.5cm}
\\
{\it
  When $G_{se} > (\ln 10)(k-2) + \ln{4}$, \( G_{mc} = \: 2G_{se} -
  \epsilon\), \( 0 \le \epsilon < \frac{1}{2k-3}, k>2 \) and $l$ is
  even, then $\sum_{p=l+1}^{\infty} N_p < 2 N_l$.}
\vspace{.5cm}
\\

To prove this lemma, we will prove two sub-lemmas.  First,

\begin{lemma}
  If $G_{se} > \left(\ln{4}\right)(k-2) + \ln{\frac{12}{5}}$, \( G_{mc} = \: 2G_{se} - \epsilon\),
  $l$ is even and \( 0 \le \epsilon < \frac{1}{2k-3} \),  then $N_{l+1} < \frac{1}{2} N_l$.
\label{fast_convergence_lemma_odd}
\end{lemma}

\begin{proof}
We will partition the
assemblies of length $l+1$ into classes corresponding to assemblies of
length $l$.  We will then show the total spurious nucleation rate of
reactions containing the assemblies in each class is at least twice as
small as the spurious nucleation rate of reactions containing its
corresponding assembly.  The class of assemblies of length $l+1$
corresponding to an assembly $B$ of length $l$ will be denoted
$\hat{B}$.

To assign the assemblies to classes, we introduce a procedure that
takes an assembly $A$ of width $k-1$ and length $l+1$, and 
then ``condenses'' its right end to yield an assembly $B$ with width $k-1$ and length $l$.
Specifically, $A$ and $B$ are identical except for the last two columns of $A$ and the last column of $B$ and if $A$ had a tile in either the ultimate or penultimate column in some particular row, then $B$ will have a tile in its last column in the same row.   Recall that for valid zig-zag assemblies, if a tile is present in a particular spot, its tile type is determined by its neighbors -- thus, we don't have to specify tile types in our condensation procedure, since there is no choice.
Formally, we say that $B = \mathbf{condensation}(A)$ if

\begin{multline*}
\forall 0 \le a < k-1,\; 0 \le b < l-1 : \; \tilde{A}(a,b)=\tilde{B}(a,b), \text{and } \\
\forall 0 \le a < k-1 :  \; \tilde{B}(a,l-1) = 0 \text{\ iff\ } \tilde{A}(a,l-1) = \tilde{A}(a,l) = 0.
\end{multline*}

Recall that $\tilde{A}$, the canonical representation of $A$, begins indexing sites at 0, so the first column has index 0 and the last ($l+1^{\text{st}}$) column has index $l$.  Also note that since $l$ is even, $A$ cannot have a double tile extending into its last column, so no double tiles are condensed.

To see that for every assembly $A$,
$\mathbf{condensation}(A)$ is bound, note first that $A$ is an
assembly, so it is bound.  
Furthermore, the connectivity graph of $B=\mathbf{condensation}(A)$ (with a vertex for each tile and an edge for each abutting pair) is just a graph-theoretic contraction of the connectivity graph of $A$ that combines any two vertices in the same row of the last two columns of $A$ (then possibly adding some extra edges).  Therefore, $B$ remains bound.
%
Thus, each $A$ of width $l+1$ is assigned to a unique, valid assembly $B$ of width
$l$.

Condensation is many-to-one, so there are many assemblies $A$ that condense onto the same smaller assembly $B$.  We assign $A$ to the class corresponding to the assembly
$\mathbf{condensation}(A)$, {\it i.e.}, the class
$$\hat{B} = \left\{ A : \mathbf{condensation}(A) = B \right\}.$$
For a given assembly $B$ of length $l$, the elements of $\hat{B}$, all of length $l+1$, 
can be created by adding $p$ tiles  ($1 \le p
\le k-2$) to the $l+1^{\text{st}}$ column of $B$, 
and then removing $h$ tiles ($0 \le h \le p-1$) from the $l^{\text{th}}$
column.

Imagine making these changes one at a time, say from top to bottom, in each row either moving or adding a tile.
For each of the $p-h$ tiles that are added to the $l+1^{\text{st}}$
column where the corresponding tiles in the $l^{\text{th}}$ column are
not removed, $p-h$ tiles are added to the assembly and no more than
$2(p-h)-1$ bonds may be formed.  For the $h$ tiles that are moved from
the $l^{\text{th}}$ to the $l+1^{\text{st}}$ column, no tiles are
added, and no more bonds can be created (some might even be lost).  Therefore, for each such assembly $A$,
$$[A]_{ss} \le e^{-(p-h)G_{mc}} e^{(2(p-h)-1)G_{se}} [B]_{ss}.$$

Let $l_A$ be the number of spurious nucleation reactions that an
assembly $A$ is a reactant of.  The rate of spurious nucleation events involving
assemblies of length $l+1$ is therefore given by:
\begin{align*}
N_{l+1}&= \sum_{\substack{A,C \in \mathcal{A}\\\text{s.t.}\; A + t\rightarrow C + t  \in R^{in}_k\\ length(A)=l+1}}
k_f[A]_{ss}e^{-G_{mc}} 
\end{align*}
We now partition this sum by summing over all smaller assemblies $B$, and then for each $A \in \hat{B}$ (recall $\hat{B} = \{A \; \text{s.t.}\;  \mathbf{condensation}(A)=B \}$) we count the spurious nucleation reactions:
\begin{align*}
=  & \sum_{\substack{B \in \mathcal{A} \\ \text{s.t.}\; length(B)=l}} \; \;
        \sum_{\substack{A \in \hat{B}, C \in \mathcal{A}\\\text{s.t.}\; A + t\rightarrow C + t  \in R^{in}_k}}k_f[A]_{ss}e^{-G_{mc}} \\
\le & \sum_{\substack{A,B \in \mathcal{A}\\\text{s.t.} \mathbf{condensation}(A) = B \\ length(B)=l}} l_A k_f[A]_{ss}e^{-G_{mc}}
\end{align*}
Partitioning $\hat{B}$ according to the number of tiles added and moved, and using our inequality for $[A]_{ss}$ in terms of $[B]_{ss}$, we have:
\begin{align*}
\le & \sum_{\substack{B \in \mathcal{A}\\ \text{s.t.}\; length(B) = l}} \; \sum_{p=1}^{k-2} {k-2 \choose p} \sum_{h=0}^{p-1} {p-1 \choose h} 
    l_A k_f[B]_{ss} e^{-(p-h)G_{mc}} e^{(2(p-h)-1)G_{se}} e^{-G_{mc}}
\end{align*}
Under the conditions of the lemma, $G_{mc} > 2G_{se} - \frac{1}{2k-3}$ so that 
\begin{align*}
< & \; \sum_{\substack{B \; s.t.\\ length(B) = l}} \sum_{p=1}^{k-2} {k-2 \choose p} \sum_{h=0}^{p-1} {p-1 \choose h} 
    l_A k_f[B]_{ss} e^{-2(p-h)G_{se}} e^{\frac{p-h}{2k-3}}e^{(2(p-h)-1)G_{se}} e^{-G_{mc}}\\
= & \; \sum_{\substack{B \; s.t.\\ length(B) = l}} \sum_{p=1}^{k-2} {k-2 \choose p} \sum_{h=0}^{p-1} {p-1 \choose h} 
   l_A k_f[B]_{ss}  e^{\frac{(p-h)}{2k-3}}e^{-G_{se}} e^{-G_{mc}}\\
= & \; \sum_{\substack{B \; s.t.\\ length(B) = l}} l_A k_f[B]_{ss} \sum_{p=1}^{k-2} {k-2 \choose p} e^{\frac{p}{2k-3}} e^{-G_{se}}
\sum_{h=0}^{p-1}  {p-1 \choose h}  e^{\frac{-h}{2k-3}} e^{-G_{mc}}
\end{align*}
Noting that the inner sums are binomial expansions of (e.g. $(1+x)^n = \sum_{i=0}^n {n \choose i} x^i$) or portions thereof, we can simplify further:
\begin{align*}
= & \; \sum_{\substack{B \; s.t.\\ length(B) = l}} l_A k_f[B]_{ss} \sum_{p=1}^{k-2} {k-2 \choose p} e^{\frac{p}{2k-3}} e^{-G_{se}}
(1 + e^{\frac{-1}{2k-3}})^{p-1} e^{-G_{mc}}. \\
\end{align*}
Since for $k>2$, $\frac{1}{2} < (1 + e^{\frac{-1}{2k-3}})^{-1} < \frac{3}{5}$,
\begin{align*}
< & \; \sum_{\substack{B \; s.t.\\ length(B) = l}} \frac{3}{5} l_A k_f[B]_{ss} \sum_{p=1}^{k-2} {k-2 \choose p} e^{\frac{p}{2k-3}} e^{-G_{se}}
(1 + e^{\frac{-1}{2k-3}})^p e^{-G_{mc}} \\
< & \; \sum_{\substack{B \; s.t.\\ length(B) = l}} \frac{3}{5} l_A k_f[B]_{ss} \sum_{p=1}^{k-2} {k-2 \choose p} e^{\frac{p}{2k-3}} 2^p 
e^{-G_{se}} e^{-G_{mc}} \\
< & \; \sum_{\substack{B \; s.t.\\ length(B) = l}} \frac{3}{5} l_A k_f[B]_{ss} \left(1 + 2e^{\frac{1}{2k-3}}\right)^{k-2} 
e^{-G_{se}}e^{-G_{mc}}.
\end{align*}
Similarly, for $k>2$, $(1 + 2 e^{\frac{1}{2k-3}}) < 4$, and $l_A \le l_B + 1$ since the longer assembly $A$ can have at most one more spurious nucleation reaction than $B$, so
\begin{align*}
\le & \; \sum_{\substack{B \; s.t.\\ length(B) = l}} \frac{3}{5}(l_B + 1)k_f[B]_{ss} e^{\ln(4)(k-2)} e^{-G_{se}} e^{-G_{mc}}\\
\le & \;\sum_{\substack{B \; s.t.\\ length(B) = l}} \frac{6}{5}l_B k_f[B]_{ss} e^{\ln(4)(k-2)} e^{-G_{se}} e^{-G_{mc}}.
\end{align*}

When $G_{se} > \ln(4)(k-2) + \ln(\frac{12}{5})$, 
\begin{align*}
< & \;\sum_{\substack{B \; s.t.\\ length(B) = l}} \frac{1}{2}l_Bk_f[B]_{ss} e^{-G_{mc}}\\
= & \;\frac{1}{2}\sum_{\substack{A + t\rightarrow B + \mathit{t}  \in R^{in}_k\\ \text{s.t.}\;length(A)=l}} k_f[A]_{ss} e^{-G_{mc}}  = \frac{1}{2} N_l.
\end{align*}

\end{proof}

The above sub-lemma takes care of the smaller odd terms, but to show
that the entire summation is bounded, we show that the smaller even terms
are also bounded.

\begin{lemma}
  If $G_{se} > \ln(10)(k-2) + \ln(4)$, \( \; G_{mc} >
  \: (2G_{se} - \frac{1}{2k-3}), k > 2 \) and $l$ is even, then $N_{l+2} <
  \frac{1}{2} N_l$.
\label{fast_convergence_lemma_even}
\end{lemma}

\begin{proof}
  The proof for this sub-lemma is similar to that for
  Lemma~\ref{fast_convergence_lemma_odd}, except that the condensation
  function is defined so that the presence of a double tile in the
  $l+1^{\text{st}}$ and $l+2^{\text{nd}}$ columns is taken into
  account.

  Here, we use a procedure that takes an assembly $A$ of width $k-1$
  and length $l+2$, and then condenses its right end to yield an
  assembly $B$ with width $k-1$ and length $l$.  Again, $A$ and $B$
  are identical except for the rightmost three columns of $A$ and the
  last column of $B$ and if $A$
  has a tile in any of the last three columns in some
  particular row, then $B$ will have a tile in its last column in the
  same row.  An added detail is that we must now consider that the
  rightmost two columns of $A$ may contain a double tile; in this
  case, the rightmost two columns of $B$ must have a double tile also.
  The double tile may either be on the top or on the bottom; without
  loss of generality, we assume it is on the bottom, since the other
  case can be treated identically.  Again, the tile types of the new
  tiles in $B$ are determined by their neighbors.  Formally, we say
  that $B = \mathbf{condensation'}(A)$ if
\begin{align*}
& \forall 0 \le a < k-1,\; 0 \le b < l-1, (a,b) \neq (k-2,l-2) : 
 \tilde{A}(a,b)=\tilde{B}(a,b), \text{ and } \\
& \forall 0 \le a < k-1 : \; \tilde{B}(a,l-1) = 0 \text{\ iff\ } 
\tilde{A}(a,l-1) = \tilde{A}(a,l) = \tilde{A}(a,l+1) = 0, \text{ and } \\
& \tilde{B} (k-2,l-2) = 0 \text{\ iff\ } \tilde{A}(k-2,l-2) = \tilde{A}(k-2,l) =\ 0.
\end{align*}

The proof that every assembly $A$ has a bound $\mathbf{condensation'}$
is virtually identical to the proof in the previous lemma.  The rest
of the proof is also similar, except that different numbers of tiles
may be removed from the $l+1^{\text{st}}$ and $l+2^{\text{nd}}$
columns.

For a given assembly $A$, creating $\tilde{A}$ from $\tilde{B}$, where
$\mathbf{condensation'}(A)=B$, requires adding $p$ tiles, $1 \le p \le
2k-3$, to the $(l+1)^{\text{st}}$ and $(l+2)^{\text{nd}}$ columns of
$B$, and then removing $h$ tiles, $1 \le h < k-1$, from the
$l^{\text{th}}$ column.

For each of the $p-h$ tiles that are added to the $(l+1)^{\text{st}}$
column and $(l+2)^{\text{nd}}$ columns where the corresponding tiles in
the $l^{\text{th}}$ column are not removed, $p-h$ tiles added to the
assembly and no more than $2(p-h)-1$ bonds may be formed.  For the $h$
tiles that are moved from the $l^{\text{th}}$ to the $(l+1)^{\text{st}}$
column or $(l+2)^{\text{nd}}$, no tiles are added, and no more bonds can
be created.

Thus, the spurious nucleation rate of these assemblies is given by:
\begin{align*}
N_{l+2}&=\sum_{\substack{A + t\rightarrow C + t  \in R^{in}_k \\ \text{s.t. } length(A)=l+2}}
k_f[A]_{ss}e^{-G_{mc}} \\  
< & \sum_{\substack{A,B \\ \text{s.t.} \mathbf{condensation'}(A) = B \\ length(B)=l}} l_A k_f[A]_{ss}e^{-G_{mc}} \\
< & \sum_{\substack{B \; s.t.\\ length(B) = l}} \sum_{p=1}^{2k-3} {2k-3 \choose p} \sum_{h=0}^{k-2} {k-2 \choose h} 
\;l_A k_f[B]_{ss} e^{-G_{mc}} e^{-(p-h)G_{mc}} e^{(2(p-h)-1)G_{se}}.
\end{align*}
When $G_{mc} > 2G_{se} - \frac{1}{2k-3}$, this similarly reduces to
\begin{align*}
< & \sum_{\substack{B \; s.t.\\ length(B) = l}} l_A k_f [B]_{ss} e^{-G_{mc}} e^{-G_{se}} 
(1+e^{\frac{-1}{2k-3}})^{k-2} (1+e^{\frac{1}{2k-3}})^{2k-3}  \\
< & \sum_{\substack{B \; s.t.\\ length(B) = l}} l_A k_f [B]_{ss} e^{-G_{mc}} e^{-G_{se}} 
\left( (1+e^{\frac{-1}{2k-3}}) (1+e^{\frac{1}{2k-3}})^2\right)^{k-2} .
\end{align*}
For $k>2$, $(1+e^{\frac{-1}{2k-3}}) (1+e^{\frac{1}{2k-3}})^2 < 10$, and thus
\begin{align*}
< & \sum_{\substack{B \; s.t.\\ length(B) = l}} 
   l_A k_f [B]_{ss} e^{-G_{mc}} e^{-G_{se}}  10^{k-2}.
\end{align*}
Therefore, when $G_{se} > \ln(10)(k-2) + \ln(4)$, and recalling that
$l_A \le l_B+1$,
\begin{align*}
< & \;\sum_{\substack{B \; s.t.\\ length(B) = l}} \frac{1}{4}(l_B+1)k_f[B]_{ss} e^{-G_{mc}}\\
< & \;\sum_{\substack{B \; s.t.\\ length(B) = l}} \frac{1}{2}l_Bk_f[B]_{ss} e^{-G_{mc}}\\
= & \;\frac{1}{2}\sum_{\substack{A + t\rightarrow B + \mathit{t}  \in R^{in}_k\\ \text{s.t.}\;length(A)=l}} k_f[A]_{ss} e^{-G_{mc}} = \frac{1}{2} N_l.
\end{align*}

\end{proof}
\vspace{4mm}
Now, we can combine Lemma~\ref{fast_convergence_lemma_odd} and Lemma~\ref{fast_convergence_lemma_even} to derive Lemma~\ref{fast_convergence_lemma}.
If $l$ is even,
\begin{align*}
\sum_{p=l+1}^\infty N_p & = N_{l+1} + N_{l+2} + N_{l+3} + N_{l+4} + \ldots \\
& < \frac{1}{2} N_l + \frac{1}{2} N_l + \frac{1}{4} N_l + \frac{1}{4} N_l + \ldots \\
& < 2 N_l.
\end{align*}

\bibliographystyle{siam}
\bibliography{journalslong,molecbib} 
\end{document}